\documentclass{aa} 
\usepackage{txfonts}
\usepackage{color}
\usepackage{graphicx}          
\usepackage{amsmath}           
\usepackage{amssymb}           
\usepackage{url}               
\usepackage{multirow}          
\usepackage{float}
\usepackage{mathabx}
\def\apjl{ApJL}

\def\apjs{ApJS}

\def\aap{A\&A}

\newcommand{\adeg}{$^{\circ}$}                   
 
\newcommand{\amin}{$^{\prime}$}                  

\newcommand{\asec}{$^{\prime \prime}$}           
\newcommand{\asecdot}[2]{\mbox{#1$\stackrel {\prime \prime}{_{\bf \cdot}}$#2}}

\newcommand{\epaqr}{EP\,Aqr}                     
\newcommand{\gapprox}{$\stackrel {>}{_{\sim}}$}  
\newcommand{\lapprox}{$\stackrel {<}{_{\sim}}$}
\newcommand{\um}{$\mu$m}                         
\begin{document} 

\title{How to disentangle geometry and mass-loss rate from AGB-star spectral energy distributions}
\subtitle{The case of EP\,Aqr
\thanks{Based on observations made with the ISO and {\it Herschel} satellites. The ISO is an ESA project with instruments funded by ESA Member States (especially the PI countries: France, Germany, the Netherlands, and the United Kingdom) and with the participation of ISAS and NASA. {\it Herschel} is an ESA space observatory with science instruments provided by European-led Principal Investigator consortia and with important participation from NASA.}}

\author{J. Wiegert \inst{1}
          \and
        M.A.T. Groenewegen \inst{1}
          \and
        A. Jorissen \inst{2}
          \and
        L. Decin \inst{3}
          \and
        T. Danilovich \inst{3}
       }

\institute{Koninklijke Sterrenwacht van Belgi{\"e}, Ringlaan 3, 1180 Brussels, Belgium\\ \email{joachimwiegert.astro@gmail.com}
\and Institut d'Astronomie et d'Astrophysique, Universit{\'e} Libre de Bruxelles, ULB, Av. F. Roosevelt 50, 1050 Bruxelles, Belgium
\and Departement Natuurkunde en Sterrenkunde, Instituut voor Sterrenkunde, KU~Leuven, Celestijnenlaan 200D, 3001 Leuven, Belgium
          }
\date{Received 26 March, 2020; accepted 24 August, 2020}

\abstract
{High-angular-resolution observations of asymptotic giant branch (AGB) stars often reveal non-spherical morphologies for the gas and dust envelopes.}
{We aim to make a pilot study to quantify the impact of different geometries (spherically symmetric, spiral-shaped, and disc-shaped) of the dust component of AGB envelopes on spectral energy distributions (SEDs), mass estimates, and subsequent mass-loss rate (MLR) estimates. We also estimate the error made on the MLR if the SED is fitted by an inappropriate geometrical model.}
{We use the three-dimensional Monte-Carlo-based radiative-transfer code RADMC-3D to simulate emission from dusty envelopes with different geometries (but fixed spatial extension). We compare these predictions with each other, and with the SED of the AGB star \epaqr\ that we use as a benchmark since its envelope is disc-like and known to harbour spiral arms, as seen in CO.}
{The SEDs involving the most massive envelopes are those for which the different geometries have the largest impact, primarily on the silicate features at 10 and 18~\um . These different shapes originate from large differences in optical depths. Massive spirals and discs appear akin to black bodies. Optically thick edge-on spirals and discs (with dust masses of $10^{-4}$ and $10^{-5}$~M$_\odot$) exhibit black-body SEDs that appear cooler than those from face-on structures and spheres of the same mass, while optically thick face-on distributions appear as warmer emission. We find that our more realistic models, combined spherical and spiral distributions, are 0.1 to 0.5 times less massive than spheres with similar SEDs. More extreme, less realistic scenarios give that spirals and discs are 0.01 to 0.05 times less massive than corresponding spheres. This means that adopting the wrong geometry for an AGB circumstellar envelope may result in a MLR that is incorrect by as much as one to two orders of magnitude when derived from SED fitting.}
{}



\maketitle


\section{Introduction}

The asymptotic giant branch (AGB) is a late evolutionary stage of low- and intermediate-mass stars (i.e. with initial masses from 0.8~M$_\odot$ to 8~M$_\odot$), just before they evolve into planetary nebulae (PNe) and white dwarfs \citep[see][for reviews]{habing1996,habing2003,hoefner2018}. Due to strong stellar winds and low surface gravities, AGB stars have high mass-loss rates (MLRs) and are surrounded by circumstellar envelopes (CSEs), inside which molecules and dust grains form. The dust is formed within a few (up to 10) stellar radii, $R_\star$, from the stellar surface (with $R_\star\sim 0.5$~au to a few au). Radiation pressure on dust continuously pushes it outwards. Drag between dust and gas drives the CSE gas outwards as well, forming an outflow of material from the star whose content enriches the interstellar medium.

Dust is an important ingredient of CSEs; however, the mechanisms governing dust formation and outflows are not well understood yet. For example, there exists much evidence that many AGB stars have companions \citep{jorissen2003,maercker2012,lagadec2015, kervella2016,jorissen2018} that are massive enough to induce discs and spirals in CSEs. Subsequently, these can significantly affect outflows \citep{theuns1993,mastrodemos1999} and, finally, the morphologies of PNe \citep[][]{jones2017}. However, spherically symmetric distributions are usually adopted by default when modelling CSEs of stars lacking detailed information on their CSE geometry. These models, which include dust radiative transfer, are used to compute spectral energy distributions \citep[SEDs;][]{elitzur2001} so that dust parameters can be constrained (e.g. dust species and grain size distributions). The knowledge of the dust parameters is essential, as they control a wide range of physical processes, for example, dust formation \citep[e.g.][and references therein]{hoefner2018}, surface chemistry that can occur on dust grains in gas-rich environments \citep[e.g.][]{herbst2005}, and radiation pressure on the dust grains controlling the outflows \citep[e.g.][]{hoefner2018,bladh2019}.

Interferometric observations now make it possible to reach the necessary milliarcsecond (mas) scales that are required to resolve CSE details such as spirals. For an AGB star at a distance of \lapprox\,$1$~kpc, 1~mas corresponds to sub-au distances, that is to say sizes of the order of the radius of an AGB star. The Atacama Large Millimeter Array (ALMA) can reach some 20\,mas to 40\,mas and the Very Large Telescope Interferometer (VLTI) can reach, at best, 2\,mas. Hence it is possible to directly observe the details of the morphology of a CSE \citep{maercker2012,kervella2014,lykou2015,paladini2017,homan2018}.

Analyses of the impact of the dust-cloud morphology on the SED have been done previously in the case of PNe and post-AGBs. For example, using direct imaging, \citet{ueta2001a} found  a toroidal structure for the post-AGB star HD~235858 (proto-PN IRAS 22272+5435). SED modelling of the central source indicates that dust grains with different sizes reside in the AGB and post-AGB shells. Similarly, \citet{murakawa2008} observed the post-AGB star IRAS 09371+1212 (`Frosty Leo'), and used non-spherical two-dimensional simulations to estimate the total CSE mass and MLR. They were able to reproduce the data with two dust species at different locations in the nebula. \citet{oppenheimer2005} published observations and radiative-transfer simulations of the dust in three proto-PNe (IRAS 17441-2411, IRAS 08005-2356, and IRAS 04296+3429). The first two are narrow-waist bipolar proto-PNe, and this extreme two-lobe structure is likely caused by a  binary companion \citep{soker2000}. However, these studies combined SED fitting and image reproduction of data where the CSE morphology was resolved.

The impact of geometry on the dusty SEDs of post-AGB stars has been known for decades \citep{vanwinckel2003}. In general, in post-AGB systems surrounded by a circumbinary disc \citep[e.g.][]{oomen2018,kamath2019,ertel2019}, the CSE contains warm dust with the IR excess starting at shorter wavelengths, typically from 1 to 3 \um .

In the case of AGB stars, \citet{jeffers2014} directly observed, for the first time, an equatorial density enhancement around an AGB star, namely IRC~+10216 (CW~Leo). They interpreted this as either a torus or a ring, and simulated the radiative transfer with these geometries. The resulting SEDs, however, appear as good as identical for both geometries. \citet{cernicharo2015} and \citet{decin2015} were able to detect spherical shells in the CSE of CW\,Leo, which were later confirmed by \citet{guelin2018}. These shells can be explained by a companion.

\citet{blum2014} analysed \textit{Spitzer}/MIPS \citep{rieke2004} spectra of a large number of Large Magellanic Cloud AGB stars. Among these stars they found that HV\,915 exhibits a disc-like signature in its spectrum, visible as CO band head emission. Normally this is found in spectra of young stellar objects and is attributed to dense and hot gas in discs. However, a specific study on the impact of morphology on the dust SEDs of AGB stars has, to the best of our knowledge, not been done yet.

The questions we address in this paper are whether it is possible to identify features in dusty SEDs that are caused by specific dust-cloud morphologies, and whether these signatures are unique to specific morphologies. We also investigate the possibility of the opposite case where different morphologies with different dust masses possibly give rise to the same SED. With this, we want to evaluate by how much CSE dust masses may be mis-estimated when adopting an incorrect dust morphology. To investigate these questions, we used the AGB star \epaqr\ as a benchmark. Its CSE has been the subject of several detailed observations \citep{nhung2019,homan2018,hoai2019,tuananh2019} which point at spiral arms embedded in a disc-like geometry, as seen nearly face-on.

This paper has the following structure. We summarise the stellar properties of \epaqr\ in Sect.~2; the dust, spatial grid, and grain properties in Sect.~3, and the simulations and statistical tools in Sect.~4. In Sect.~5 we present our results which are discussed in Sect.~6 and summarised in Sect.~7.

\section{Properties of EP Aqr} 
\label{Sect:properties}

EP Aquarii (\epaqr , HD\,207076, HIP\,107516, Gaia DR2 2673831344664664320) is an oxygen-rich star on the AGB. Table~\ref{starprops} summarises its properties. There have been several recent investigations of the CSE around \epaqr\ \citep{nhung2015,homan2018,nhung2019,hoai2019,tuananh2019}, most of them of the CO lines with ALMA.

\begin{table}
    \caption{\epaqr\ observed and adopted properties.}
    \label{starprops}
    \begin{center}
    \begin{tabular}{rl}
\hline\hline
\noalign{\smallskip}
    \multicolumn{2}{c}{Observed properties} \\
\noalign{\smallskip}
\hline
\noalign{\smallskip}
    RA$^a$                    &  21$^{\rm h}$46$^{\rm m}$31.847$^{\rm s}$ \\
    DEC$^a$                   & -02\adeg 12\amin \asecdot{45}{902} \\
    $l$                       & 54\adeg 2\\
    $b$                       & $-39$\adeg 3\\
    Spectral type$^b$         & M7-III \\
    Distance$^{c{\rm ,\,1}}$  & $113.6 \pm 8.2$\,pc \\
    Luminosity$^d$            & 4828~L$_\odot$ \\ 
    Radius$^{a}$              & 87.5${+5.0}\atop{-10.7}$~R$_\odot$ \\
    Effective temperature$^e$ & 3236\,K \\ 
    Mass$^e$                  & 1.7~M$_\odot$ \\
    $E_{B-V}$ (ISM)$^f$       & 0.058 mag\\
\noalign{\smallskip}
\hline
\noalign{\smallskip}
    \multicolumn{2}{c}{Adopted properties for MARCS} \\
\noalign{\smallskip}
\hline
\noalign{\smallskip}
    Effective temperature     & 3200\,K \\
    Mass                      & 1~M$_\odot$ \\
    Metallicity ([Fe/H])      & 0.0 dex \\
    Surface gravity, $\log g$ & 0.5 dex \\
\noalign{\smallskip}
\hline
    \end{tabular}
    \end{center}
    \begin{list}{}{}
    \item[Notes.]
     $^1$ We note that there are more recent parallaxes from Gaia DR2, however, at the start of this project we noted large error bars for data on bright stars \citep[e.g.][]{lindegren2018,drimmel2019} and opted to use the older Hipparcos parallax for this study.
    \item[References.]
    (a) \citet{gaia2018}
    (b) \citet{keenan1989}
    (c) \citet{leeuwen2007}
    (d) \citet{winters2003}
    (e) \citet{dumm1998}
    (f) \citet{gontcharov2012}
    \end{list}
\end{table}

\citet{tuananh2019} found jets extending from 25 to 1000\,au. In the CO emission, \citet{homan2018} found a vertically confined spiral extending up to $\sim10$\asec\ ($\sim\,1000$~au) from \epaqr . The one armed-spiral is visible between 1\asec\ and 5\asec\ from the star where it exhibits two revolutions. This spiral is both flat and seen nearly face-on (with an inclination between 4\adeg\ and 18\adeg ). At \asecdot{0}{5} from the star, a local void is observed in SiO emission and a bridge of gas between the central star and the void in CO emission, both indicating the presence of a (substellar) companion with an upper mass limit of 0.1~M$_\odot$. The existence of a spiral around this star makes it a good candidate for our tests exploring spectral features in SEDs specifically due to spirals or discs.

\citet{nhung2015} measured a MLR for \epaqr\ of $\sim~1.2\times 10^{-7}$~M$_\odot\,$yr$^{-1}$ and a terminal outflow velocity of 10 to 11~km~s$^{-1}$ (see their figure~12), in agreement with more recent values from \citet{hoai2019}. These authors find a MLR of $(1.6 \pm 0.4) \times 10^{-7}$~M$_\odot$~yr$^{-1}$, an outflow velocity of 10 to 11~km~s$^{-1}$ from the stellar poles, and an outflow velocity of 2~km~s$^{-1}$ at the stellar equator.

\subsection{Stellar SED model and observations}

We used the MARCS\footnote{\url{http://marcs.astro.uu.se/}} grid of stellar model atmospheres \citep{gustafsson2008} to extract the photospheric spectrum that we used as  input for RADMC-3D\footnote{\url{http://www.ita.uni-heidelberg.de/~dullemond/software/radmc-3d/}}, a three-dimensional radiative-transfer code to model the SED under various geometries \citep[][and Sect.~\ref{Sect:simulations}]{dullemond2012}. RADMC-3D is written in Fortran and handles arbitrary dust spatial distributions, with any dust species, with any number of stars, and with the possibility of adding molecular lines. The input MARCS synthetic spectrum covers a wavelength range up to 20~\um , after which we extrapolated the spectrum with a black body (BB). 

We searched the MARCS grid for models with solar metallicity, an effective temperature $T_{\rm eff}$ of 3200\,K, and with current stellar masses of 1 or 2~M$_\odot$, which are close to the temperature and mass of \epaqr , namely 3236~K and 1.7~M$_\odot$, respectively \citep{dumm1998}. The properties of the adopted MARCS model are listed in Table~\ref{starprops} together with observed properties. We chose a model with a mass of 1~M$_\odot$ since such models better fit the observed photometry in the  visual and ultraviolet (V and UV) domains, that is the domains which we normalised the MARCS model SED to, as listed in Table~\ref{allfluxes}. For comparison we plotted the photometry on top of this MARCS model in the last frame of Fig\,\ref{largestarseds}. Not listed in Table\,\ref{allfluxes}, but included in this study, are spectra from the {\it Infrared Space Observatory} Short Wavelength Spectrograph (SWS) as extracted by \citet{sloan2003}\footnote{\url{https://users.physics.unc.edu/~gcsloan/library/swsatlas/aot1.html}} and previously analysed by \citet{heras2005}, and \textit{Herschel}-PACS \citep{pilbratt2010,poglitsch2010} spectra from \citet{nicolaes2018}.

\begin{table}
    \caption{Photometry of \epaqr . See also Fig.\,\ref{largestarseds}}
    \label{allfluxes}
    \begin{center}
    \begin{tabular}{ccr}
\hline\hline
\noalign{\smallskip}
    $\lambda_{\rm eff}$ & $S_{\nu}$ & Photometry \\
    (\um)               & (Jy)      & Reference  \\
\noalign{\smallskip}
\hline
\noalign{\smallskip}
   0.528  & $   15.10   \pm    1.51^a$  & Hipparcos (1) \\ 
   0.674  & $   85.55   \pm    0.59 $   & Gaia (2) \\   
   1.26   & $ 2093.6    \pm  391.8 $    & 2MASS J (3) \\
   1.60   & $ 3626.6    \pm  925.0$     & 2MASS H (3) \\
   2.16   & $ 3214.5    \pm 1135.6$     & 2MASS K (3) \\
   3.6    & $ 1450.31   \pm 145.03^a$   & Johnson L (1) \\
   3.8    & $ 1411      \pm  141^a$     & UKIRT L' (4) \\ 
   4.8    & $  802      \pm   80^a$     & UKIRT M  (4) \\ 
  12      & $  432.65   \pm   43.27^a$  & IRAS (1) \\ 
  18      & $  382.40   \pm   38.24^a$  & Akari L18W (1) \\ 
  25      & $  228.30   \pm   22.83^a$  & IRAS (1) \\ 
  70      & $   22.3    \pm    2.2^a$   & \textit{Herschel}-PACS (5) \\ 
 100      & $   16.4    \pm    1.6^a$   & IRAS (6) \\ 
 100      & $   10.0    \pm    1.0^a$   & \textit{Herschel}-PACS (5) \\ 
 100      & $   15.0    \pm    1.5^a$   & \textit{Herschel}-PACS (6) \\ 
 160      & $    4.7    \pm    0.5^a$   & \textit{Herschel}-PACS (6) \\ 
 160      & $    3.7    \pm    0.4^a$   & \textit{Herschel}-PACS (5) \\ 
1300      & $    0.016  \pm    0.001^b$ & IRAM (7) \\ 
1300      & $    0.0170 \pm    0.0003^b$& ALMA (8) \\ 
1300      & $    0.0178 \pm    0.0002$  & ALMA (9) \\ 
2600      & $    0.0045 \pm    0.0004^b$& IRAM (7) \\ 
2600      & $    0.0049 \pm    0.0001$  & ALMA (9) \\ 
\noalign{\smallskip}
\hline
    \end{tabular}
    \end{center}
    \begin{list}{}{}
    \item[Notes.] 
$^a$ Assumed 10\% error bar. $^b$ Error bar based on background RMS.
    \item[References.] 
(1) \citet{mcdonald2017}
(2) \citet{gaia2018}
(3) \citet{cutri2003}
(4) \citet{fouque1992}
(5) \citet{nicolaes2018}
(6) \citet{ramosmedina2018}
(7) \citet{winters2007}
(8) \citet{homan2018}
(9) \citet{hoai2019}
    \end{list}
\end{table}

We would like to mention a few caveats concerning stellar variability and reddening with how this stellar SED model was implemented in RADMC-3D. Both V-UV, and the infrared to far-infrared (IR and FIR) wavelength regimes are simultaneously sensitive to the surrounding dust (from reddening or dust emission, respectively) and variability of the star itself. However, the IR and FIR is more sensitive to dust emission than to variability. We do not take variability into account here since we focus on the effects from the dust itself for a pulsation-averaged \epaqr . However, to consider variability we would need to re-simulate all dust morphologies for a set of stellar SEDs that correspond to different phases of the star's variability. This would not in any case be feasible since such a set of models would be static, and not take time-dependent propagation of variable radiation fields into account. 

Reddening becomes an issue when the dust envelope is optically thick. Our input stellar SED model is normalised to the observed flux densities at V-UV wavelengths. Thus we initially do not take reddening into account when we search for a stellar SED model. Furthermore, \citet{heras2005} found that the real dust envelope of \epaqr\ is optically thin, which is also evident later in this study where we compare simulated dust SEDs with observed data. However, it is also shown later in this study that effects from reddening is significant for the more massive dust envelopes. If we were to search for correct dust envelope models, that fit the data, we would take reddening into account when we search for stellar SED models. However, our focus in this study is to better understand the effects of dust emission by comparing SEDs from different dust morphologies and with different total dust masses, including the more massive and optically thick dust envelopes. As such we use one stellar SED model and focus on the effects on the SEDs in the IR and FIR regimes by varying the dust models only. It is also worth noting that the interstellar reddening computed from \citet{gontcharov2012} map is considered to be negligible (Table~\ref{starprops}).

\section{Description of dust properties} 

\subsection{Dust composition and optical properties}
\label{sect:dustcomp}

In Table\,\ref{dustprops} we summarise the dust properties used for our simulations. Since \epaqr\ is an O-rich AGB star, we opt to include common dust species for such stars. Two of the most common Si-bearing species are fosterite (Mg$_2$SiO$_4$) and enstatite \citep[MgSiO$_3$;][]{hoefner2018}, and these are often found in combination with alumina (Al$_2$O$_3$) and fayalite \citep[Fe$_2$SiO$_4$, which forms further out from the star;][]{gobrecht2016,millar2016}. We opt to primarily include 99\%\ amorphous Mg$_2$SiO$_4$ and 1\%\ crystalline Fe$_2$SiO$_4$. This was chosen for simplicity, due to the clear silicate features in the ISO-SWS spectrum, and for example \citet{jones2014} showed that, for O-rich AGB stars in the Large Magellanic Cloud, dust consists primarily of amorphous Si species. However, in our comparison with observational data, we also used another composition mixture containing 9\%\ amorphous Al$_2$O$_3$ dust, plus 90\% Mg$_2$SiO$_4$ and 1\% Fe$_2$SiO$_4$. This composition was estimated from the flux density ratio at the 10~\um\ silicate feature and at the 13~\um\ Al-feature in the ISO-SWS spectrum. It is consistent with the abundance of Al$_2$O$_3$ in the AGB stars observed by \citet{jones2014} which have MLR similar to \epaqr . For a study of the actual dust content of \epaqr , we refer to the analysis of ISO-SWS spectra of AGB stars by \citet{heras2005}. They fitted five different dust species to the spectrum and, in their figure~2, a good match is obtained with $\sim 9$\% Al$_2$O$_3$, $\sim 80$\% of a mixture of silicates, and $\lesssim\,10$\% of iron-rich dust.

\begin{table}
    \caption{List of simulation parameters.}
    \label{dustprops}
    \begin{center}
    \begin{tabular}{rl}
\hline\hline
\noalign{\smallskip}
    \multicolumn{2}{c}{Grain properties$^1$} \\
\noalign{\smallskip}
\hline
\noalign{\smallskip}
    Density$^a$          & 3\,g\,cm$^{-3}$ \\ 
    Size$^b$             & 0.15 - 0.27~\um $^2$ \\ 
\noalign{\smallskip}
    Species and          & Mg$_2$SiO$_4$, 99\%  (alt. 90\%) \\ 
    abundances$^{b,c,d}$ & Fe$_2$SiO$_4$, 1\%   (alt. 1\%) \\
                         & Al$_2$O$_3$, 0\%  (alt. 9\%) \\
\noalign{\smallskip}
Maximum albedo, $\eta_{\rm Max}$ & Mg$_2$SiO$_4$: 0.99 ($\lambda = 0.67$\um ) \\
                                 & Fe$_2$SiO$_4$: 0.99 ($\lambda = 0.68$ \um )\\
                                 & Al$_2$O$_3$:   0.73 ($\lambda = 0.27$ \um )\\
\noalign{\smallskip}
    Average scattering   & Mg$_2$SiO$_4$: $g = 0.59$ \\
    angle, $g = \langle \cos{\theta} \rangle$, & Fe$_2$SiO$_4$: $g = 0.53$ \\
    at maximum albedo    & Al$_2$O$_3$: $g = 0.87$ \\
\noalign{\smallskip}
\hline
\noalign{\smallskip}
    \multicolumn{2}{c}{Dust cloud properties$^1$} \\
\noalign{\smallskip}
\hline
\noalign{\smallskip}
    Total dust masses$^4$         & $10^{-8}$~M$_\odot$ to $10^{-4}$~M$_\odot$ \\
    Radii                         & 5 to 5000 au \\
    Density profile, $\rho(r)$    & $\propto\,r^{-2}$ \\
    Spiral inter-arm              & \multirow{2}{*}{250 au} \\
    distance, $2\,\pi b$          & \\
    Spiral \&\ disc thickness$^5$ & 9 au$^5$ \\
\noalign{\smallskip}
\hline
\noalign{\smallskip}
    \multicolumn{2}{c}{Other simulation parameters} \\
\noalign{\smallskip}
\hline
\noalign{\smallskip}
    $N$ thermal photons        & $10^7$ \\
    $N$ scattering photons$^6$ & $10^6$ \\ 
    $N$ spectral photons$^6$   & $10^6$ \\ 
    Scattering type$^e$        & \textit{Henyey-Greenstein} approximation \\
    Wavelength range           & 0.1 to 4000~\um\ \\
    $N$ wavelength bins$^7$    & 520 \\ 
    Spatial grid               & 4 level-refined octree \\ 
\noalign{\smallskip}
\hline
    \end{tabular}
    \end{center}
    \begin{list}{}{}
    \item[Notes.]
    $^1$ Adopted value(s).
    $^2$ Gaussian distribution centred on 0.20~\um .
    $^3$ $\langle \cos{\theta} \rangle$ is the average of the cosine of the scattering angle.
    $^4$ In five logarithmic steps (see text for details).
    $^5$ Spiral cross section is a square because of the spatial grid. The disc thickness increases when the distance to the star is larger than 128~au, see Sect.\,\ref{sect:models} for details.
    $^6$ Per wavelength bin.
    $^7$ Logarithmic range plus photometry wavelengths.
    \item[References.] 
    (a) \citet{moromartin2013}
    (b) \citet{hoefner2018}
    (c) \citet{gobrecht2016}
    (d) \citet{jones2014}
    (e) \citet{henyey1941}
    \end{list}
\end{table}

We extracted the refractive indices of the considered dust species from the \textit{Databases of Dust Optical Properties} from the \textit{Astrophysical Laboratory Group of the AIU Jena}\footnote{\url{https://www.astro.uni-jena.de/Laboratory/Database/databases.html}}. The original references for these optical properties are, for fosterite \citet{jaeger2003}, for fayalite \citet{fabian2001}, and for alumina \citet{begemann1997}. These were used to compute the mass-absorption and scattering coefficients for input into RADMC-3D. A version of the \citet{bohren1983} Mie code, originally written in Fortran, and rewritten in Python by C. Dullemond in 2017 for inclusion in RADMC-3D, was used for this. This code assumes spherical grains with a Gaussian grain size distribution and a logarithmic size spread. The Gaussian spread reduces artefacts caused by a too-narrow, unrealistic size range of spherical grains, such as resonances and sharp features in the spectra. Recent results (\citealt{norris2012,ohnaka2016,ohnaka2017}, and summarised in review by \citealt{hoefner2018}) state that grains of sizes 0.1 to 0.5~\um\ have been detected in the vicinity of AGB stars. We thus adopt relatively small and compact grains with a size distribution centred on 0.2~\um\ (in effect 11 different grain sizes between 0.15 and 0.27~\um\ in logarithmic bins). For simplicity we kept the grain size constant with distance to the star so that we may have the same extinction coefficients throughout the dust distribution. We note that Mg$_2$SiO$_4$ and Fe$_2$SiO$_4$ grains of these sizes have albedos close to one at wavelengths around 0.7~\um\ (see Table\,\ref{dustprops}). This will result in very strong contributions from scattered light to the dust SED in the full optical range, as we will discuss later (Sect.\,\ref{sect:results}).

The RADMC-3D code includes several methods for simulating scattering. Since the Bohren \&\ Huffman Mie code computes scattering angles, we compared results with no scattering, isotropic scattering, and anisotropic \textit{Henyey-Greenstein} approximated scattering \citep{henyey1941}. Simulations with no scattering are obviously faster, but also unrealistic since most of the scattering occurs at short wavelengths where most of the heating of the dust also happens (through absorption). For example, with Rayleigh scattering, we may assume that most scattering occurs at wavelengths shorter than 10~\um\ for grains with the size of 0.2~\um . There were only insignificant differences in the SED when comparing the two different scattering modes. The computational time was comparable and even shorter with anisotropic scattering. Thus we chose to use anisotropic \textit{Henyey-Greenstein} scattering in our simulations. Because of heavy computational effort for the densest geometries, we reduced the scattering tolerance in the simulations. By default RADMC-3D destroys scattered photons when the optical thickness reaches 30 and we reduced this limit to five as recommended by the author of RADMC-3D in situations of computationally heavy simulations. We also compared SEDs of both settings and found no noticeable differences.

The choice of the grain mass density of 3~g~cm$^{-3}$ is based on findings that interplanetary dust particles in the solar system have densities ranging between 1 and 3~g~cm$^{-3}$ depending on their origin and population \citep{moromartin2013}. Since we are interested in the impact  on the SED of the CSE morphology, we kept the grain density constant for all models.

\subsection{Dust morphology and spatial grid description}
\label{sect:models}

Our focus in this study is the spiral-shaped morphology, however, we include a disc for comparisons. As such we limited the simulations to four CSE morphologies: a sphere, a spiral, a disc, and a combined sphere-spiral envelope. With these we include the extremes with no spherical component, and intermediate cases of a spiral embedded in a sphere.

The gas spiral that was found by \citet{homan2018} around \epaqr\ is the basis for the spiral models. This spiral was observed between the radial distances $\sim 1$\asec\ and 5\asec\ from the central star and follows an Archimedean spiral with 2 to 2.5 revolutions. At the distance of \epaqr , the observed spiral thus covers the region extending from 100 to 500~au from the star. The extension of our spiral model goes from 5 to 5000~au with a radial inter-arm distance of 250~au (corresponding to 2 revolutions within 500~au, as observed). The inner radius of 5~au corresponds to $\sim 10\,R_\star$, which is approximately the inner limit for iron-rich dust formation (e.g. Fe$_2$MgSiO$_4$, see e.g. \citealt{hoefner2018}). Because of numerical constraints, we set this as the inner limit for all dust species. However, we must note that we miss some hot dust emission by not including dust within $\sim 2$ to $10\,R_\star$, with the effect that we predict a marginally colder dust SED than if this dust had been included. On the other hand, we used this radial limit for all models so the comparisons between them are valid. The outer limit of 5000~au for all dust species is based on Herschel/PACS images of \epaqr , revealing that the dust shell extends out to about 0.04~pc (or $\sim 8250$~au) at both 70 and 160~$\mu$m \citep[see Fig.~1 of][]{cox2012}. 5000~au is again a compromise since it is numerically difficult to simulate three-dimensional models of larger sizes. However, this is an acceptable compromise since the dust emission from such a large distance from the star has only marginal effects on the resulting SED.

\begin{figure}
    \centering
        \includegraphics[width=80mm]{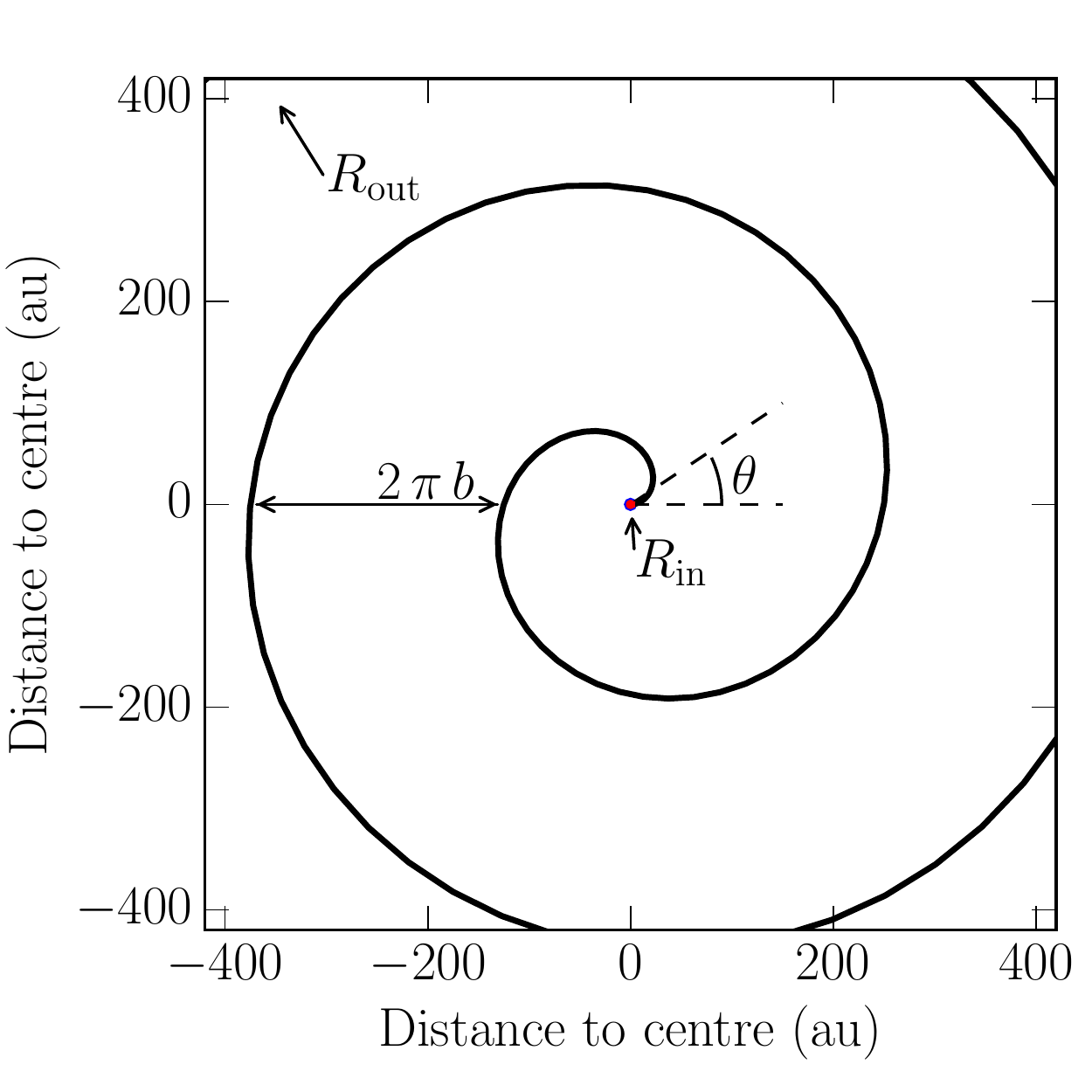}
    \caption{Schematic overview of spiral parameters. The inner radius for the dust is indicated by $R_{\rm in}$ at 5~au, while the outer radius $R_{\rm out}$ is beyond the limits of this image. The same radii are adopted as limits for the disc and spherical geometries. The parameters $\theta$ and $b$ are defined in Eq.\,\ref{eqspiral}.}
        \label{figspiral}
\end{figure}

An Archimedean spiral is described as
\begin{equation}
r = a + b\,\theta,
\label{eqspiral}
\end{equation}
where $r$ is the distance to the centre of the star, $\theta$ is the running azimuthal angle, $a$ is the radius at $\theta = 0$, and $b$ controls the width between each revolution of the spiral (see Fig.\,\ref{figspiral} for a schematic overview of these parameters). 

To simply implement spirals in an octree refined cubic grid we chose to keep the vertical extent equal to each arm's horizontal width. We limit the vertical extent to 90\%\ of the radius of the empty region close to the star (defined by $R_{\rm in}$), that is 4.5~au. These choices are based on limitations of a cubic grid and the assumption that the spiral of \epaqr\ is confined to a plane, as was seen by \citet{homan2018}.

An octree grid is a cubic grid where each refinement level equally divides grid-cells of a higher level grid. In three-dimensions, this means that a grid-cell is divided into eight refined cells for each level of refinement. As we will further discuss below, we applied four levels of refinement in the area around the spiral arms, inside the disc, or in the area around the inner edge of the sphere, in the case of spherical distributions. The side of the base cell (i.e. 32~au) is thus 16 times larger than that of the finest, fourth-level grid cell.

In the spiral model, we increasingly refined the spatial grid around the centre of the spiral, with the first refinement being one base cube size from the centre of the spiral. All maximally refined cells of the spirals were assigned dust densities computed from a pre-defined total dust mass and the radial power law $\rho\propto r^{-2}$ (Table\,\ref{dustprops}). The finest grid cells of 2.0~au were chosen as a compromise between best grid resolution at the inner radius $R_{\rm in}$ and reasonable computing time. Because we implemented a cubic grid we also obtained a square-shaped cross section through the spiral arm.

In the disc model, we applied a coarser grid than in the spiral model due to computational constraints. Since the grid directly affects the dust envelope morphology, the thickness of the disc is 9~au only when $r < 128$~au (as stated in Table\,\ref{dustprops}). The disc thickness then increases to 13~au within $128 \le r < 256$~au, 17~au within $256 \le r < 512$~au, and finally to 25~au outside 512~au. The emission at wavelengths larger than 100~\um\ from the dust located farther away than 1000~au in these thick discs represents less than 10\%\ of the emission from thinner discs. We then conclude that this way of modelling discs is an acceptable compromise between accuracy and computational convenience.

The grid for the spherical models was maximally refined in the area around the inner radius of the sphere, but could generally be coarser than what we used for spirals and discs. The grid cells are of the same sizes as for the other morphologies and the refinement steps are instead located at 10, 19, 35, and 67~au.

The combined sphere-spiral models include the same spiral as described above inside a spherical dust model. The spiral component of the spatial grid is thus combined with the spherical dust model spatial grid. From the literature we estimate that an Archimedean spiral with our constraints should be around 1\%\ of the total dust mass of the envelope. For example, \citet{kim2019} suggest that spirals are $\sim 2$ times the density of the inter-arm regions, while \citet{guelin2018} found shells in the envelope of CW\,Leo that have $\sim 3$ times the density of the intershell regions, which, with our spiral model, gives a spiral mass that is $\sim 0.3$ to 0.6\%\ of the total dust mass. A more massive spiral-example can be seen in figures 5 and 9 of \citet{chen2019} where we see that the densities of their spirals are around $\sim 10^2$ to $\sim 10^3$ (and even as high as $10^4$) times that of the inter-arm regions. This would give a spiral mass of $\sim$10\%\ to $\sim$50\%\ of the total dust mass in our model. Evidently there exist a wide range of possible ratios. However, in general the ratio is small so we limit our masses to the lower values 0.5\%\ and 5\%\ in the spiral component, and 99.5\%\ and 95\%\ in the spherical component, respectively.

The discs and spirals were primarily simulated at face-on (0\adeg ) and edge-on (90\adeg ) inclinations. The inclination can have a particularly important impact on optically thick cases, as will be discussed in Sect.\,6.4 where we also include simulations at inclinations of 25\adeg , 50\adeg , 60\adeg , 70\adeg , 80\adeg , 85\adeg , and 90\adeg . For example, \citet{ueta2001b} stress the importance of the inclination angle in the case of modelling emission of PNe and how different morphologies are hard to disentangle under certain inclinations.

All geometrical models follow the same radial density power law $\rho~\propto~r^{-2}$, where a density normalisation factor was used to obtain the input total dust masses. For each geometry, we used five different total dust masses in logarithmic steps, from $10^{-8}$ to $10^{-4}$~M$_\odot$. To make the comparison with other studies easier, we estimate the total MLRs corresponding to these dust masses. For this, we assumed a gas-to-dust ratio of 100, which is a value commonly derived from observations in various environments. For example, \citet{meixner2004} found a ratio of 75 in the proto-PN HD~56126, and \citet{danilovich2015} found ratios close to 100 for several AGB stars (see their table~5 where $h$ is a function of the gas-to-dust ratio).

A radial density power law proportional to $r^{-2}$ and a constant outflow velocity $v(r) = v_\infty$ imply a constant MLR. We assumed $v(r) = 11$\,km\,s$^{-1}$, which is the terminal outflow velocity as found by \citet{nhung2015} for \epaqr . The total (gas and dust) mass density of a spherical CSE is then given by the law of mass conservation,
\begin{equation}
\rho_{\rm tot}(r) \approx\, \frac{\dot M_{\rm tot}}{4\,\pi\,r^2\,v_\infty}.
\label{csedensity}
\end{equation}

Since we assumed a constant MLR and a constant terminal velocity, we can simply calculate the amount of mass distributed between 5 and 5000~au. Including a gas-to-dust ratio of 100 one obtains
\begin{equation}
M_{\rm dust}(5\,{\rm to}\,5000\,{\rm au}) = \frac{\dot M_{\rm tot}\; [{\rm M_\odot\; yr^{-1}}]}{v_\infty [{\rm au\; yr}^{-1}]} \times \frac{4995\;[{\rm au}]}{100},
\label{mlrmass}
\end{equation}
where $v_\infty [{\rm au\; yr}^{-1}] \approx 2.32$.

With the above assumed terminal velocity and spherical CSE geometry, the dust mass to total MLR correspondence is
\begin{equation}
\begin{cases}
10^{-4}~{\rm M}_\odot \Leftrightarrow\, 4.64\times 10^{-6 }&{\rm M_\odot\,yr}^{-1}\\
10^{-8}~{\rm M}_\odot \Leftrightarrow\, 4.64\times 10^{-10}&{\rm M_\odot\,yr}^{-1}.
\end{cases}
\label{largemlrmass}
\end{equation}
Thus we can state that the observed \epaqr\ MLR of $1.2\times 10^{-7}$~M$_\odot\,$yr$^{-1}$ \citep{nhung2015} corresponds to $\sim 2.6\times 10^{-6}$~M$_\odot$ of dust within 5 to 5000~au.

\section{Simulations and statistics} 
\label{Sect:simulations}

We used RADMC-3D \citep{dullemond2012} to simulate SEDs for the CSE geometries described in Sect.~\ref{sect:models}. We primarily ran five simulations for each geometry, one for each dust mass given in Sect.~\ref{sect:models}. All the other parameters are listed in Table~\ref{dustprops}. The total simulation time using 20 to 30 cores for each SED is about $\sim\,12$ hours for those models with the highest densities.

The output data of RADMC-3D can be either in the form of SEDs over the whole input wavelengths grid, images at arbritrarily chosen wavelengths within the wavelength grid, or spectra at certain wavelength-sub-ranges. We mainly used the SED outputs for this study. It is possible to retrieve the SED as it would be seen by an observer, or the SED resulting from the dust component only. To more easily distinguish differences in the dust emission, we used the dust SED in much of the analysis here, but we also noted when these differences would be detectable for an observer. We approximated a general gauge for this purpose from the average error of ISO-SWS spectra which is $\sigma_{\rm ISO} \approx 3.6$~Jy. Whenever we only plot the dust SED output from RADMC-3D, we denote it as the {\it dust SED}.

To gauge the differences, or similarities, between different simulated SEDs we used
\begin{equation}
\chi^2_{\rm model} = \frac{1}{N} \sum_{\nu }\,\frac{\left[ S_\nu ({\rm model}\ 1) - S_\nu ({\rm model}\ 2) \right]^2}{S_\nu ({\rm model}\ 2)},
\label{eqchimodel}
\end{equation}
which is a reduced Pearson-$\chi^2$ test ($N$ is the number of wavelength bins). In the above relation, $S_\nu ($model$)$ is the predicted flux density. 

For each geometry and inclination, we searched for the  model best fitting the observed SED of \epaqr . This gave us actual examples of how different the MLRs may be, depending on the adopted geometry. The best fit was found by searching for the $\chi^2_{\rm red}$ minimum, with $\chi^2_{\rm red}$ defined as 
\begin{equation}
\chi^2_{\rm red} = \frac{1}{N-1}\,\sum_{\nu }\,\left[ \frac{S_\nu ({\rm obs}) - S_\nu ({\rm model})}{\sigma_\nu} \right]^2,
\label{eqchidata}
\end{equation}
where $S_\nu ($obs$)$ are \epaqr\ observed flux densities, either derived for the photometric passbands listed in Table\,\ref{allfluxes}, or from ISO-SWS and \textit{Herschel}-PACS spectra \citep[][respectively]{sloan2003,nicolaes2018}. The $\sigma_\nu$ parameter is the corresponding error bar at frequency $\nu$ of the observed SED. The reduced $\chi^2$ has $N-1$ degrees of freedom, where $N$ is the number of wavelength bins, since we constrain the total dust mass from this minimisation process. We convolved our simulated SEDs with photometric filters downloaded from the SVO Filter Profile Service\footnote{\url{http://svo2.cab.inta-csic.es/theory/fps/}} \citep{rodrigo2012,rodrigo2013} when comparing with observed photometry.

\section{Results} 
\label{sect:results}

\begin{figure*}
    \centering
        \includegraphics[width=157mm]{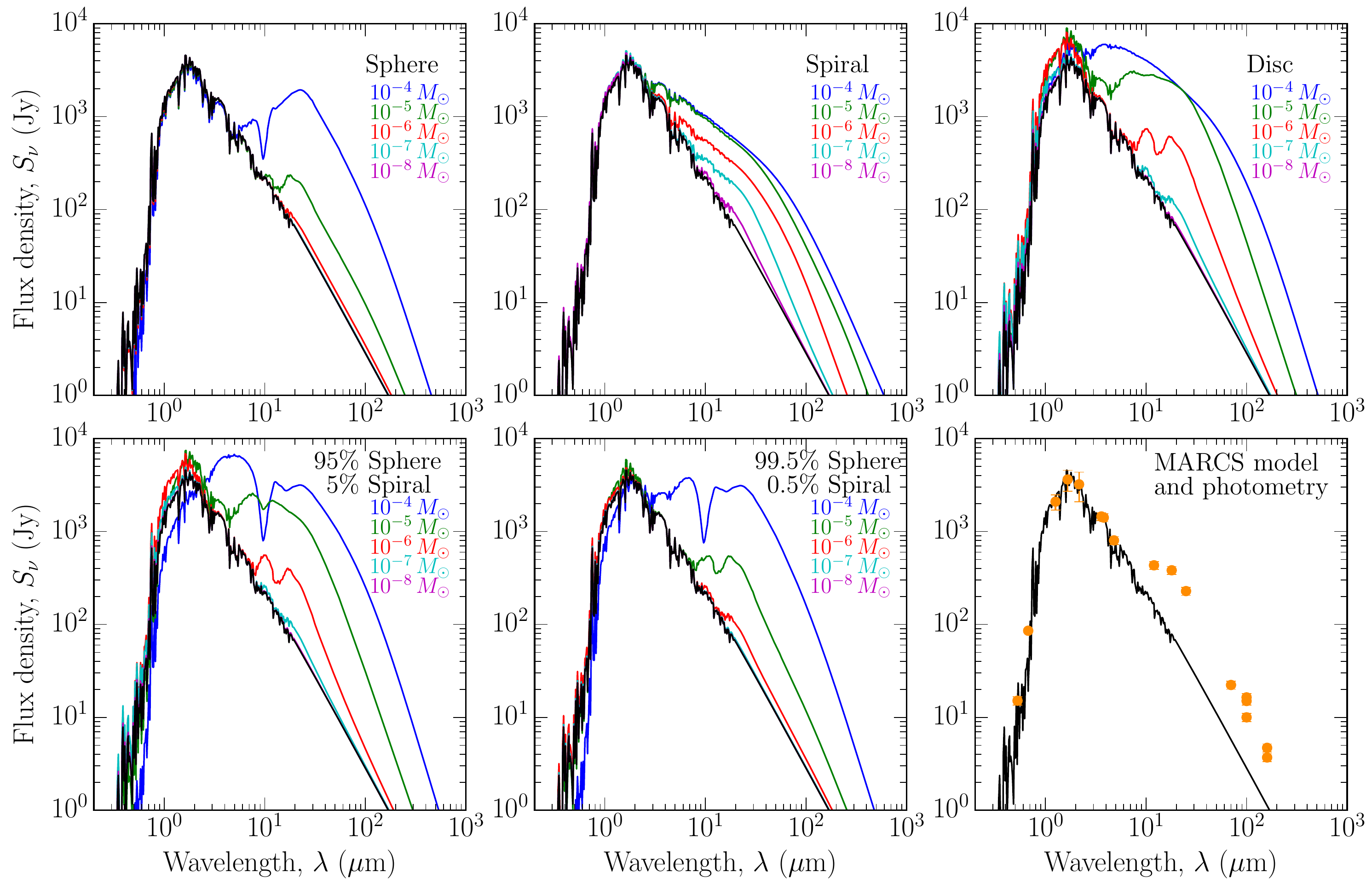}
    \caption{All face-on, 5 to 5000~au simulated SEDs, and the stellar SED MARCS model (in black). The dust masses are $10^{-4}$ (blue), $10^{-5}$ (green), $10^{-6}$ (red), $10^{-7}$ (cyan), and $10^{-8}$~M$_\odot$ (magenta). Top row shows SEDs from CSE with 100\%\ sphere, spiral and disc, respectively, while the second row shows combined sphere-spiral model SEDs and a comarison between the photometry listed in Table\,\ref{allfluxes} (orange data points) and the MARCS model.}
        \label{largestarseds}
\end{figure*}

In Fig.\,\ref{largestarseds} we show the modelled SEDs with the \epaqr\ SED model, with only face-on inclinations for the spiral and disc models (i.e. spiral or disc lies in the plane of the sky, $i = 0$\adeg). Here we can see significant differences in the simulated spectra, particularly the strong silicate features at 10 and 18~\um\ for the spheres. The most massive sphere has an absorption feature at 10~\um , while the less massive spheres have emission features at 10 and 18~\um\ instead, and the most massive spirals and discs have no spectral features at these wavelengths. We also see that the most massive models differ significantly from the \epaqr\ SED model at wavelengths shorter than 1~\um . This is expected since these models contain much more dust than the real CSE of \epaqr\ actually contains. This is also not a problem since we are using this star with a known spiral as a test object for our models, and some models will of course differ.

We see that discs are associated with SEDs peaking at shorter wavelengths than the spheres, even though they have the same inner radius. Also for post-AGB stars is a warmer SED temperature indicative of a disc caused by a binary central star \citep{vanwinckel2003}. This is not conclusive, however, since also a sphere with a smaller inner radius will have a warmer SED temperature than our models.

In Table\,\ref{opticalthickness} we list the radii along the line-of-sight (LOS) within which the dust cloud optical thickness is larger than unity at any considered wavelength, assuming edge-on geometry for spirals and discs (i.e. inclination angle of 90\adeg ). Maximum optical thickness is obtained with our extinction coefficients at $\lambda = 0.38$~\um . It is, in effect, a measure of approximately how deep we see into each geometry. It can be seen that only spherical dust shells with masses less than $10^{-6}$~M$_\odot$ and dusty discs with $10^{-8}$~M$_\odot$ are (marginally) optically thin. Thus for a given dust mass, dusty spheres are optically thinner than discs and spirals, which is easily understandable since in the disc and spiral cases, the same amount of dust occupies a smaller volume.

\begin{table}
    \caption{List of radii outside which the dust distributions are optically thin for all considered wavelengths. Maximum optical thickness is obtained with our extinction coefficients at $\lambda = 0.38$~\um . For the spiral and disc models, an inclination angle of 90\adeg\ is adopted (i.e. edge-on) and no spherical component is included.}
    \label{opticalthickness}
    \begin{center}
    \begin{tabular}{cccc}
\hline\hline
\noalign{\smallskip}
    Mass    & Sphere      & Spiral      & Disc        \\
(M$_\odot$) & radius (au) & radius (au) & radius (au) \\
            & at $\tau_\nu\,\sim\,1$ & at $\tau_\nu\,\sim\,1$ &  at $\tau_\nu\,\sim\,1$  \\
\noalign{\smallskip}
\hline
\noalign{\smallskip}
    $10^{-4}$ &  655 & 5000 & 4590 \\
    $10^{-5}$ &   73 & 4990 & 2650 \\
    $10^{-6}$ &    8 & 4940 &  505 \\
    $10^{-7}$ & $<5$ & 4410 &   55 \\
    $10^{-8}$ & $<5$ & 2170 &    6 \\
\noalign{\smallskip}
\hline
    \end{tabular}
    \end{center}
\end{table}

\begin{figure*}
    \centering
        \includegraphics[width=157mm]{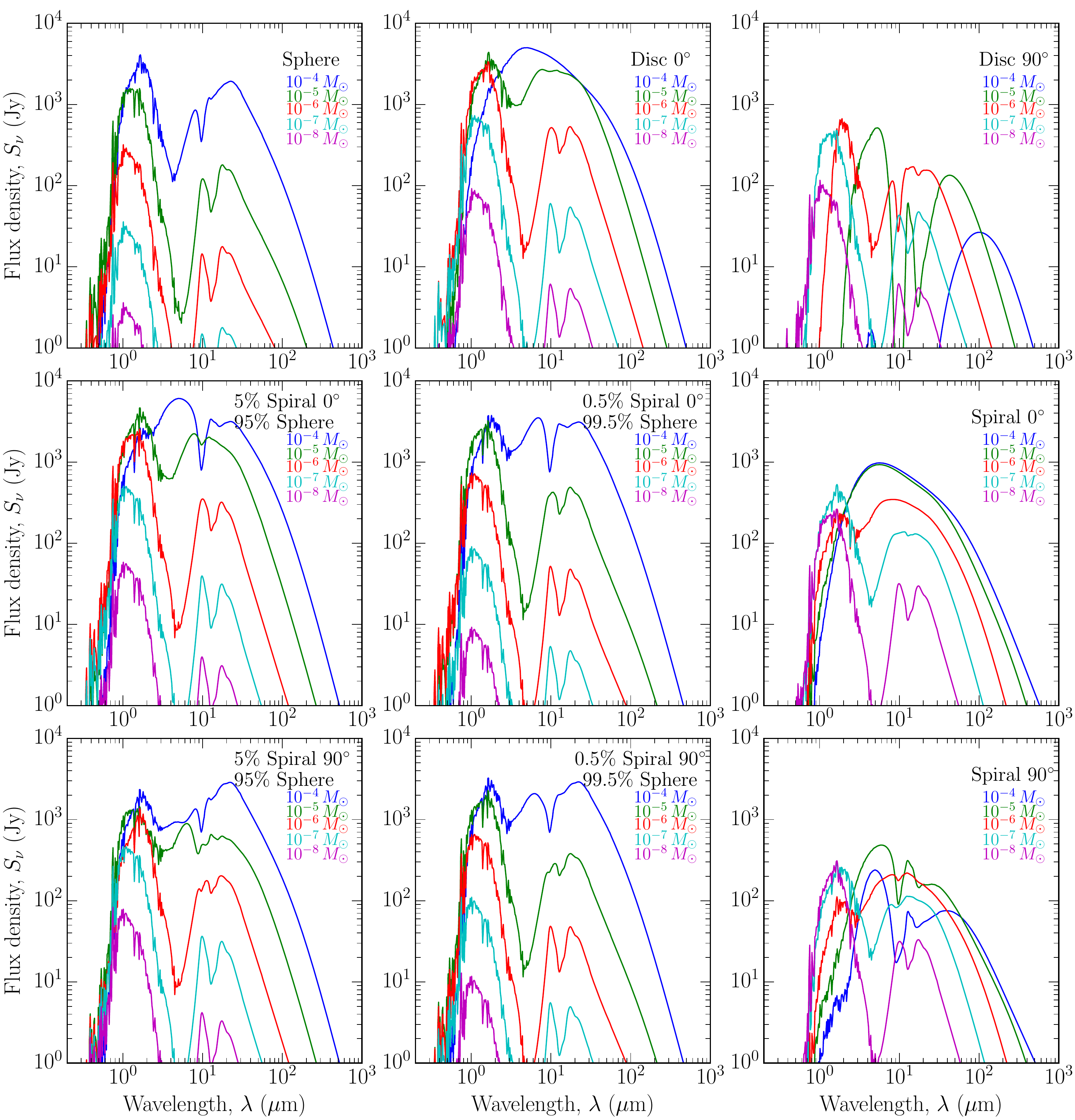}
    \caption{All 5 to 5000~au simulated dust SEDs -- without the stellar SED model -- of dust masses $10^{-4}$ (blue), $10^{-5}$ (green), $10^{-6}$ (red), $10^{-7}$ (cyan), and $10^{-8}$~M$_\odot$ (magenta). The dust cloud morphology and inclination are indicated in the top right courner of each panel. The long wavelength portion of the SEDs ($> 2$~\um ) are primarily dust heat emission while the short wavelength portion ($< 2$~\um ) are scattered light on dust from the star.}
        \label{largeseds}
\end{figure*}

This difference in optical depth impacts the emission or absorption nature of the silicate features. The effects of optical thickness are more apparent in Fig.~\ref{largeseds} where we only show the dust contribution to the SED (for both face-on and edge-on models) without the stellar SED. Here we see that silicate features in spherical models change from emission to absorption going from optically thin to optically thick dust distributions. Similarly, in the face-on spirals and discs, the silicate features change from emission to absorption, and even become featureless. Because of the smaller absorption, stars with face-on spirals and discs appear more luminous than those with only spherical envelopes.

The optically thick spirals and discs seen edge-on exhibit a very strong silicate absorption feature at 10~\um . Furthermore, the more massive edge-on discs appear as cold BBs but they are also primarily shaped by the very strong silicate absorption. In the most massive edge-on cases we only receive emission from the outer parts of the dust distribution. This results in both apparently colder and fainter SEDs. In Sect.~\ref{Sect:inclination}, we investigate in a more quantitative manner the impact of inclination.

Inclination and optical thickness also result in non-intuitive effects for the scattered light portion of the dust SED. As mentioned in Sect.~\ref{sect:dustcomp}, our dust has high albedo and this results in significant contributions of scattered light in all face-on or optically thin cases. The scattered light is visible as a peak at wavelengths around 0.3 to 5~\um\ in Fig.~\ref{largeseds}. In the optically thin cases of $10^{-7}$~M$_\odot$ and $10^{-8}$~M$_\odot$, the flux density of the scattered peak does not change much when observed face-on or edge-on, as may be seen in Fig.~\ref{largeseds} from the comparison of the `Disc 0\adeg ' and `Disc 90\adeg ' cases. However, in the optically thick spirals and discs (with higher dust masses) seen edge-on, the scattered light is significantly reduced, while it remains strong in the face-on situations in agreement with the fact that the scattering angle parameter $g$ is close to one, which corresponds to mostly forward scattering. Moreover, in the edge-on situations with large dust masses, the central parts of the discs and spirals are masked behind significant amounts of optically thick dust along the LOS. The optically thick nature of these situations is revealed by the fact that the SED now closely resembles a black body. In those cases, we only observe the outer rim of the dust distribution (see Table~\ref{opticalthickness}) whereas  in the face-on cases, the colder outer dust, the warmer inner dust, and the light scattered on the inner dust grains all contribute to the observed light.

We discuss the differences between the model SEDs in more detail in Sect.~\ref{Sect:discussion}, and mention a few details here. If we use the error bar of the ISO-SWS spectrum (on average $\sigma_{\rm ISO} \approx\,3.6$~Jy) and assume that dust contribution is detectable when its excess is more than $3\,\sigma_{\rm ISO}$ above the stellar SED, we find that spirals and discs of masses \gapprox$10^{-7}$~M$_\odot$ and spheres of masses $> 10^{-6}$~M$_\odot$ are detectable. These masses of 100\%\ spirals or discs (i.e. no spherical dust distribution included) correspond to MLRs of $4.6\times 10^{-9}$ and $4.6\times 10^{-8}$~M$_\odot\,$yr$^{-1}$, respectively. It is worth noting that we are considering mas scales and would need interferometers to resolve the spiral (since a spiral width of $\sim\,$9~au is 90~mas at $\sim\,100$~pc). However, an outer radius of 5000~au is large enough to be resolved in many observations (some 44\arcsec\ at 114~pc). The highest angular resolutions of ALMA are some 20 to 40\,mas and, as an example in the optical, the VLTI can reach some 2\,mas.

\section{Discussion} 

Here we first discuss the possibility of deducing the CSE geometry from the dusty SED (Sect.~\ref{Sect:discussion}). Then in Sect.~\ref{Sect:masscorrection}, we show how to estimate the amount of dust mass missed by assuming a spherical geometry and build a mass-correction term to approximately correct for this error. Third in Sect.~\ref{Sect:inclination}, we compare SEDs of different inclinations for the most massive distributions where the strongest impact is expected. Finally in Sect.~\ref{comparisonphotometry}, we use best-fits of each model to the observed data of \epaqr\ to test the mass correction term of Sect.~\ref{Sect:masscorrection}.

\subsection{Simulated spectral features}
\label{Sect:discussion}

\begin{figure}
    \centering
        \includegraphics[width=80mm]{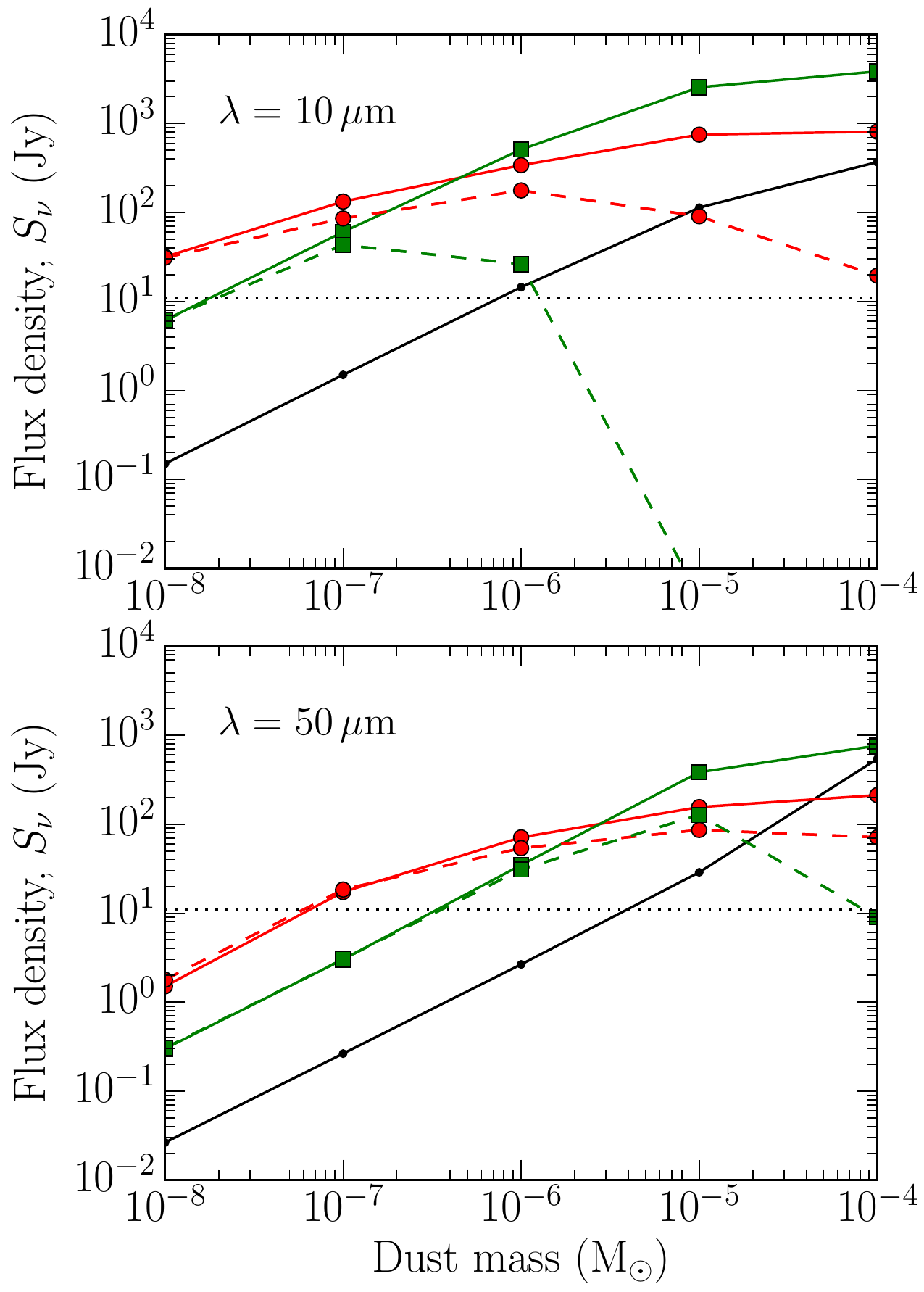}
    \caption{Comparisons of dust flux densities from Fig.\,\ref{largeseds} at 10~\um\ ({\it top} panel) and 50~\um\ ({\it bottom} panel) for spheres (black with dots), spirals (red with circles) and discs (green with squares) without a spherical component, and as seen face-on (solid lines) or edge-on (dashed lines). The horizontal black dotted line shows the flux density threshold of $3\,\sigma_{\rm ISO} \approx 10.8$~Jy.}
        \label{figspectralfeatures}
\end{figure}

The dust SEDs are shown in Fig.\,\ref{largeseds}, and dust emission at 10 and 50~\um\ from 100\%\ spirals and discs is compared with spheres in Fig.\,\ref{figspectralfeatures} to visualise differences and similarities between SEDs. Since the more massive edge-on discs are optically thick, and appear as featureless SEDs with colder black-body temperatures and less flux at wavelengths shorter than $\sim 50$~\um , their SEDs are more akin to less massive spheres with larger radii. This is most visible in Fig.\,\ref{figspectralfeatures} for SEDs from edge-on discs of masses $10^{-4}$ and $10^{-5}$~M$_\odot$, which decrease in flux density to undetectable levels at these wavelengths (below the $3\,\sigma_{\rm ISO} \approx\,10.8$~Jy), since the emission peak is at longer wavelengths.

Disc SEDs with a dust mass of $10^{-8}$~M$_\odot$ exhibit only negligible differences when compared to SEDs of spheres with dust masses $< 10^{-6}$~M$_\odot$. However, this dust excess is less than $3\,\sigma_{\rm ISO}$ above the stellar SED and would be difficult to detect in observations (compare with the horizontal dotted line in Fig.\,\ref{figspectralfeatures}). Also, the strong absorption at 10~\um\ of the edge-on disc with a dust mass of $10^{-4}$~M$_\odot$, visible in Fig.\,\ref{largeseds}, would not be detectable in reality since it would be overwhelmed by the stellar SED.

All geometrical models have the same spectral appearance when they are optically thin with silicate features at 10 and 18~\um . For the optically thick cases, all spherical geometries exhibit a strong absorption at 10~\um\ (for dust masses larger than  $10^{-5}$~M$_\odot$), while the other geometries, as mentioned above, rather show a black-body-like spectral shape when optically thick. Therefore, the lack of silicate features at 10 and/or 18~\um\ could be an indication of non-spherical dust cloud geometries since these have higher optical thickness than corresponding spherical geometries with the same dust mass (see Table\,\ref{opticalthickness}). 

To demonstrate these effects, we fitted black body SEDs to the simulated dust SEDs' thermal emission at wavelengths $> 2$~\um\ (only dust emission without stellar SED, and excluding the combined sphere-spiral geometry). Since our simulated and observed data are expressed as flux density per unit frequency and as a function of wavelength, we use the Planck function as expressed per unit frequency,
\begin{equation}
B_\nu (\lambda , T_{\rm dust}) = \frac{2 h c}{\lambda^3} \times \left[ \exp\left( \frac{h c}{k_{\rm B} T_{\rm dust} \lambda} \right) - 1 \right]^{-1},
\label{planckfunceq}
\end{equation}
where $\lambda~=~c~/~\nu$, $h$ is the Planck's constant, $k_{\rm B}$ is Boltzmann's constant, $c$ is the speed of light in vacuum, and $T_{\rm dust}$ is the dust temperature. For this reason, the maximum flux density per unit frequency as a function of wavelength, as listed in Table\,\ref{largesedbbtable}, follows the frequency dependant Wien's displacement law.

\begin{table}
    \caption{Fitted dust BB-temperatures and wavelengths, $\lambda_{\rm BB}$, that correspond to the frequency of the peak of the BB-function $B_\nu (\lambda , T_{\rm dust})$ for spherical models, and edge-on spirals and discs.}
    \label{largesedbbtable}
    \begin{center}
    \begin{tabular}{llccc}
\hline\hline
\noalign{\smallskip}
    Mass        & Model & $T_{\rm dust}$ & $\lambda_{\rm BB}$  & $R_{\rm dust}$ \\
    (M$_\odot$) &        & (K)      & (\um ) & (au)      \\
\noalign{\smallskip}
\hline
\noalign{\smallskip}
\multirow{3}{*}{$10^{-4}$} & Sphere & 227 &  22   &  152 \\
                           & Spiral & 133 &  38   &  573 \\
                           & Disc   &  50 & 102   & 6590$^a$ \\
\noalign{\smallskip}
\multirow{3}{*}{$10^{-5}$} & Sphere & 287 &  18   &   85 \\
                           & Spiral & 629 &   8   &   12$^b$ \\ 
                           & Disc   & 120 &  43   &  747 \\
\noalign{\smallskip}
\multirow{3}{*}{$10^{-6}$} & Sphere & 287 &  18   &   85 \\
                           & Spiral & 411 &  12   &   34 \\
                           & Disc   & 340 &  15   &   55 \\
\noalign{\smallskip}
\multirow{3}{*}{$10^{-7}$} & Sphere & 287 &  18   &   85 \\
                           & Spiral & 411 &  12   &   34 \\
                           & Disc   & 298 &  17   &   80$^b$ \\ 
\noalign{\smallskip}
\multirow{3}{*}{$10^{-8}$} & Sphere & 287 &  18   &   85 \\
                           & Spiral & 293 &  17   &   80 \\
                           & Disc   & 519 &  10   &   19$^b$ \\ 
\noalign{\smallskip}
\hline
    \end{tabular}
    \end{center}
    \begin{list}{}{}
    \item[Notes.]
    $^a$ Dust BB-radius may appear as $>5000$~au even though there is no dust there since the dust we do see is colder than it would have been in an optically thin model.
    $^b$ In these examples the BB-fit to the simulated SED was complicated by the presence of strong silicate features.
    \end{list}
\end{table}

To compute BB-radii ($R_{\rm dust}$), we refer to \citet{lamers1999} who state that
\begin{equation}
R_{\rm dust} = \frac{R_\star}{2}\,\left( \frac{T_\star}{T_{\rm dust}} \right)^{(4+s)/2},
\label{bbradiuseq}
\end{equation}
where $s~=~1$ for optically thin dust. $R_{\rm dust}$ then gives the edge of the dust condensation region, which we take to be the distance between star and dust. The stellar properties (radius $R_\star$ and effective temperature $T_\star$) are those listed in Table~\ref{starprops}. This way we obtain BB temperatures that correspond to the thermal emission of the dust and can more easily see how the thermal emission, as seen edge-on, moves to longer wavelengths when the dust is optically thick (more massive dust envelopes).

In Table\,\ref{largesedbbtable}, we list the fitted dust BB-radii of spheres and edge-on geometries, and in Table\,\ref{opticalthickness} we list radii along the LOS where the dust envelope becomes optically thick as seen edge-on. The sphere BB-temperatures are generally close to 290\,K, while the edge-on geometries have significantly colder dust BB-temperatures and appear as optically thin at much larger distances from the star. For example, the optically thin SEDs peak at wavelengths around 10 to 20~\um , while the more massive edge-on spirals and discs SEDs peak at wavelengths around 50 to 100~\um . A general conclusion could be that SEDs with no discernable features and peaking at longer wavelengths can originate from disc-like distributions that are optically thick and observed close to edge-on. However, we found that SEDs that are peaking at shorter wavelengths can also originate from disc-like geometries as seen more face-on rather than edge-on, as also concluded by \citet{vanwinckel2003}. Furthermore, inclusion of larger grains is also known to result in colder emission due to differences in the absorption coefficients as compared to small grains \citep[see e.g.][]{miyake1993,wolf2003,krivov2008}.

\subsection{Morphology-mass correction term}
\label{Sect:masscorrection}

To compensate for the ignorance of the dust morphology, we constructed a dust mass correction term by searching for similar SEDs from different morphologies. We did this by using Eq.\,\ref{eqchimodel} and retaining those SEDs that fall within the $\chi_{\rm model}^2$-limit of $\chi_{\rm model}^2~<~15$ when compared to SEDs of the purely spherical morphology. This is a large range since we have SEDs from a wide and sparse dust mass range, rather than a narrow and well-filled dust mass range. The $\chi_{\rm model}^2$-limit was chosen as a compromise between the number of similar SEDs (for each set of models) and the quality of the similarities.

The mass corrections $C_M$ for the different geometries are defined as
\begin{eqnarray}
C_M({\rm geometry}) = \frac{M_{\rm geometry}}{M_{\rm sphere}},
\label{eqmasscorr}
\end{eqnarray}
where the subscript {\it geometry} can be exchanged to any dust geometry. With this correction factor, an observed dust mass, deduced from a SED under the assumption of spherical geometry, can be transformed into a corresponding spiral, sphere-spiral, or disc dust mass through multiplication by the $C_M$ factor.

\begin{figure}
    \centering
        \includegraphics[width=80mm]{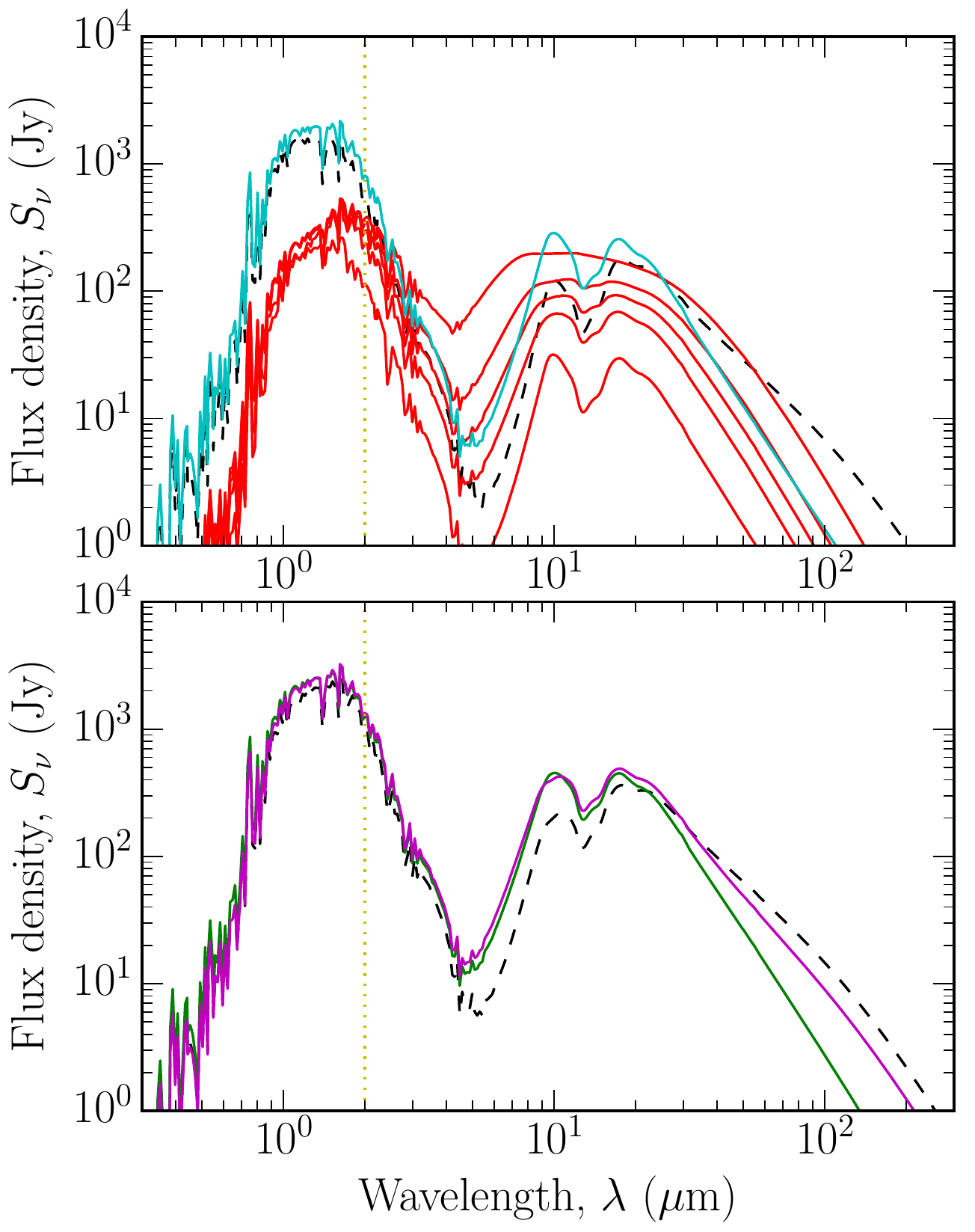}
    \caption{Comparisons of different `worst' cases of $\chi_{\rm model}^2~<~15$, for face-on spirals and discs, as listed in Table~\ref{largemassrelationtable}. The {\it top} panel shows dust SEDs corresponding to the spherical dust mass of $1~\times~10^{-5}~$M$_\odot$ (black dashed line), i.e. face-on 100\%\ spiral of mass $2~\times~10^{-7}~$M$_\odot$ and a few additional SEDs down to $1~\times~10^{-8}~$M$_\odot$ (red lines), and a face-on 5\%\ spiral with the dust mass $8~\times~10^{-7}~$M$_\odot$ (cyan line). The {\it bottom} panel shows dust SEDs corresponding to the spherical dust mass of $2~\times~10^{-5}~$M$_\odot$ (black dashed line), i.e. a face-on disc with the dust mass $8~\times~10^{-7}~$M$_\odot$ (green line), and a face-on 0.5\%\ spiral with the dust mass $1~\times~10^{-5}~$M$_\odot$ (magenta line). The vertical yellow dotted line marks the wavelength limit of 2~\um .}
        \label{chisquare50}
\end{figure}

We list spiral and disc masses, corresponding sphere masses, and subsequent $C_M$ factors (i.e. mass corrections) for each SED that fulfil $\chi_{\rm model}^2~<~15$ in Table\,\ref{largemassrelationtable}. In Fig.\,\ref{chisquare50}, we show examples of two spherical SEDs compared with face-on oriented disc and spiral SEDs. This figure showcases the bluntness of our $\chi_{\rm model}^2$-limit and the meaning of `quality' of the similarities. For example, in the case of face-on 100\%\ spirals, we see that the SEDs with the highest allowed $\chi_{\rm model}^2$ exhibit no silicate features at 10 and 18~\um\ due to optical thickness, in constrast to the other models with the highest allowed $\chi_{\rm model}^2$. It should be noted that our $\chi_{\rm model}^2$-limit was partly chosen so that it would include 100\%\ spiral SEDs that exhibit these silicate features, and along a range of dust masses until the silicate features are no longer visible. Thus, we included additional SEDs here for the $C_M$ factor of the 100\%\ spirals, than for the other envelope morphologies.

\begin{table*}
    \caption{Mass corrections for spirals, discs, and sphere-spiral combinations, corresponding sphere masses, mass corrections, and $\chi^2_{\rm model}$ numbers, as given by Eq.\ref{eqchimodel}, for all models that fulfils $\chi_{\rm model}^2~<~15$. The left half contains numbers for the face-on geometries and the right half contains numbers for the edge-on geometries.}
    \label{largemassrelationtable}
    \begin{center}
    \begin{tabular}{cccc|cccc}
\multicolumn{4}{c}{Face-on inclinations}&\multicolumn{4}{c}{Edge-on inclinations}\\
\hline\hline
\noalign{\smallskip}
$M_{\rm spiral}$&$M_{\rm sphere}$&$C_M$&$\chi^2_{\rm model}$& $M_{\rm spiral}$&$M_{\rm sphere}$&$C_M$&$\chi^2_{\rm model}$ \\
(M$_\odot$)     &(M$_\odot$)     &     &                    & (M$_\odot$)     &(M$_\odot$)     &     &                     \\
\noalign{\smallskip}
\hline
\noalign{\smallskip}
 $2\times 10^{-7}$ & $1\times 10^{-5}$ & 0.02 &  13.6   &   $1\times 10^{-7}$ & $7\times 10^{-6}$ & 0.01 &   6.7 \\ 
 $1\times 10^{-7}$ & $8\times 10^{-6}$ & 0.01 &   7.2   &   $1\times 10^{-8}$ & $2\times 10^{-6}$ & 0.005&   1.9 \\ 
 $9\times 10^{-8}$ & $8\times 10^{-6}$ & 0.01 &   6.6   & \multicolumn{4}{c}{ \dots } \\
 $8\times 10^{-8}$ & $7\times 10^{-6}$ & 0.01 &   6.0   & \multicolumn{4}{c}{ \dots } \\
 $7\times 10^{-8}$ & $7\times 10^{-6}$ & 0.01 &   5.4   & \multicolumn{4}{c}{ \dots } \\
 $6\times 10^{-8}$ & $6\times 10^{-6}$ & 0.01 &   4.7   & \multicolumn{4}{c}{ \dots } \\
 $5\times 10^{-8}$ & $5\times 10^{-6}$ & 0.01 &   4.2   & \multicolumn{4}{c}{ \dots } \\
 $4\times 10^{-8}$ & $5\times 10^{-6}$ & 0.01 &   3.6   & \multicolumn{4}{c}{ \dots } \\
 $3\times 10^{-8}$ & $4\times 10^{-6}$ & 0.01 &   3.0   & \multicolumn{4}{c}{ \dots } \\
 $2\times 10^{-8}$ & $3\times 10^{-6}$ & 0.01 &   2.3   & \multicolumn{4}{c}{ \dots } \\
 $1\times 10^{-8}$ & $1\times 10^{-6}$ & 0.01 &   1.5   & \multicolumn{4}{c}{ \dots } \\
\noalign{\smallskip}
\hline
\noalign{\smallskip}
$M_{\rm disc}$&$M_{\rm sphere}$&$C_M$&$\chi^2_{\rm model}$   &   $M_{\rm disc}$&$M_{\rm sphere}$&$C_M$&$\chi^2_{\rm model}$ \\
(M$_\odot$)   &(M$_\odot$)     &     &                       &   (M$_\odot$)   &(M$_\odot$)     &     &                     \\
\noalign{\smallskip}
\hline
\noalign{\smallskip}
 $8\times 10^{-7}$ & $2\times 10^{-5}$ & 0.04 &  14.8    &   $1\times 10^{-6}$ & $1\times 10^{-5}$ & 0.10 &  10.4 \\ 
 $7\times 10^{-7}$ & $2\times 10^{-5}$ & 0.04 &  14.6    &   $9\times 10^{-7}$ & $1\times 10^{-5}$ & 0.09 &   8.7 \\ 
 $6\times 10^{-7}$ & $1\times 10^{-5}$ & 0.06 &  14.7    &   $8\times 10^{-7}$ & $1\times 10^{-5}$ & 0.08 &   7.3 \\ 
 $1\times 10^{-7}$ & $2\times 10^{-6}$ & 0.05 &   1.6    &   $1\times 10^{-7}$ & $2\times 10^{-6}$ & 0.05 &   1.3 \\ 
 $1\times 10^{-8}$ & $2\times 10^{-7}$ & 0.05 &   0.2    &   $1\times 10^{-8}$ & $3\times 10^{-7}$ & 0.03 &   0.2 \\ 
\noalign{\smallskip}
\hline
\noalign{\smallskip}
$M_{\rm 5\% spiral}$&$M_{\rm sphere}$&$C_M$&$\chi^2_{\rm model}$   &   $M_{\rm 5\% spiral}$&$M_{\rm sphere}$&$C_M$&$\chi^2_{\rm model}$ \\
(M$_\odot$)$^a$      &(M$_\odot$)     &     &                      &   (M$_\odot$)$^a$      &(M$_\odot$)     &     &                    \\
\noalign{\smallskip}
\hline
\noalign{\smallskip}
 $8\times 10^{-7}$ & $1\times 10^{-5}$ & 0.08 &   9.7   &   $1\times 10^{-6}$ & $1\times 10^{-5}$ & 0.10 &   7.2 \\
 $1\times 10^{-7}$ & $1\times 10^{-6}$ & 0.10 &   1.3   &   $1\times 10^{-7}$ & $1\times 10^{-6}$ & 0.10 &   1.4 \\
 $1\times 10^{-8}$ & $1\times 10^{-7}$ & 0.10 &   0.1   &   $1\times 10^{-8}$ & $1\times 10^{-7}$ & 0.10 &   0.1 \\
\noalign{\smallskip}
\hline
\noalign{\smallskip}
$M_{\rm 0.5\% spiral}$&$M_{\rm sphere}$&$C_M$&$\chi^2_{\rm model}$   &  $M_{\rm 0.5\% spiral}$&$M_{\rm sphere}$&$C_M$&$\chi^2_{\rm model}$ \\
(M$_\odot$)$^a$       &(M$_\odot$)     &     &                       &  (M$_\odot$)$^a$       &(M$_\odot$)     &     &                     \\
\noalign{\smallskip}
\hline
\noalign{\smallskip}
 $1\times 10^{-5}$ & $2\times 10^{-5}$ & 0.50 &   9.7    &  $1\times 10^{-5}$ & $2\times 10^{-5}$ & 0.50 &   4.2 \\
 $1\times 10^{-6}$ & $2\times 10^{-6}$ & 0.50 &   0.7    &  $1\times 10^{-6}$ & $3\times 10^{-6}$ & 0.33 &   0.7 \\
 $1\times 10^{-7}$ & $2\times 10^{-7}$ & 0.50 &   0.1    &  $1\times 10^{-7}$ & $3\times 10^{-7}$ & 0.33 &   0.1 \\
 $1\times 10^{-8}$ & $2\times 10^{-8}$ & 0.50 &   0.01   &  $1\times 10^{-8}$ & $3\times 10^{-8}$ & 0.33 &   0.01\\
\noalign{\smallskip}
\hline
    \end{tabular}
    \end{center}
    \begin{list}{}{}
    \item[Notes.] $^a$ Mass refers here to total dust mass (sphere and spiral).
    \end{list}
\end{table*}

\begin{figure}
    \centering
        \includegraphics[width=80mm]{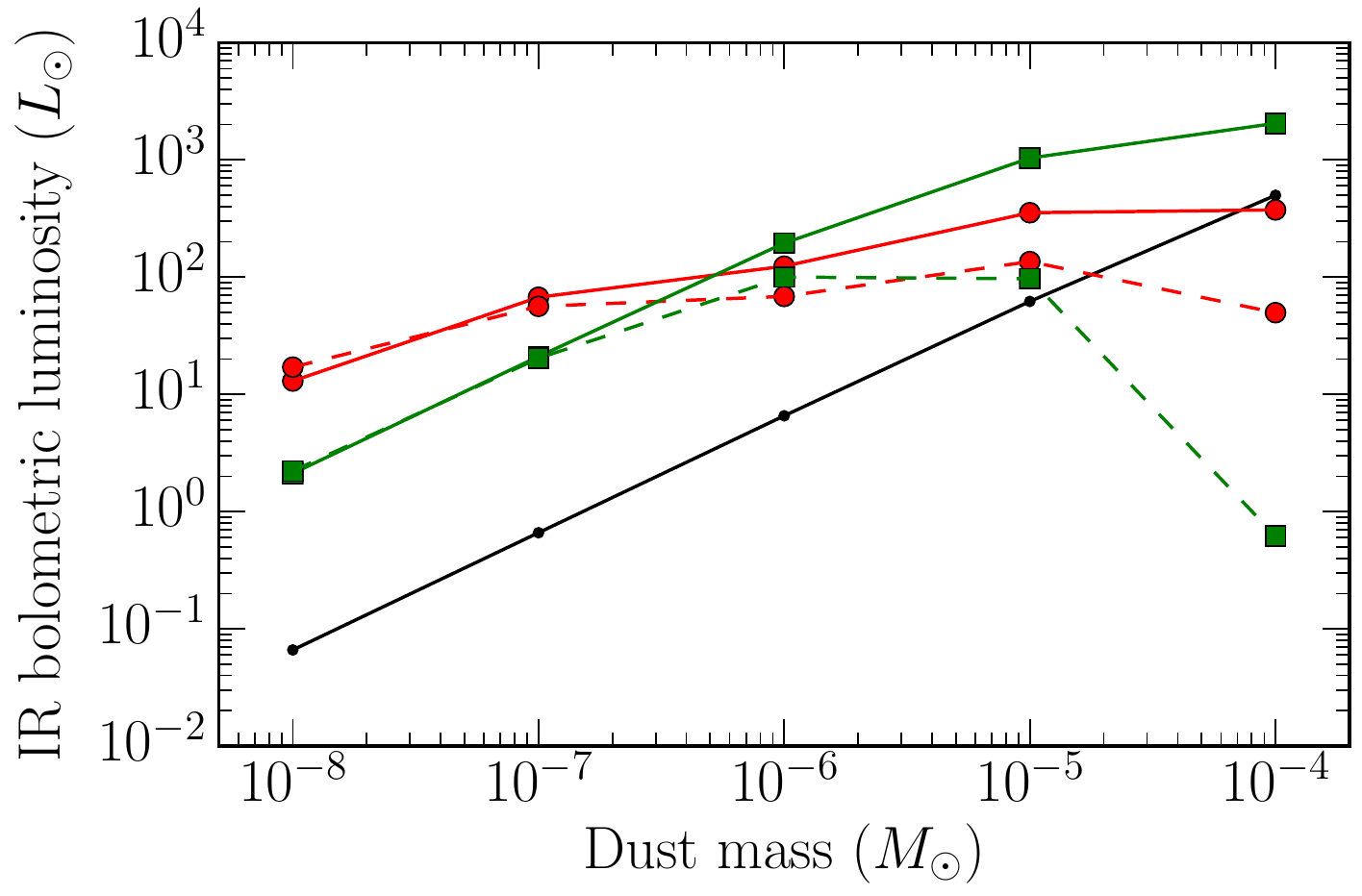}
    \caption{Dust IR luminosities for spherical distributions (black), and spirals (red dots) and discs (green squares) without a spherical component. Solid lines are face-on distributions and dashed lines are edge-on distributions.}
        \label{largelummass}
\end{figure}

We obtain, in the face-on cases, regular trends for the $C_M$ factor with growing sphere dust mass, while for most edge-on cases we obtain $C_M$ factors over a wide range. To better understand the differences of the $C_M$ factors from the edge-on geometries, we plot the bolometric dust IR luminosity ($\lambda\,>\,2$~\um ) for each dust geometry in Fig.~\ref{largelummass} (based on SEDs shown in Fig.~\ref{largeseds}). For the spheres, the IR luminosity increases linearly (in a log-log plot) with mass, which is expected for mostly optically thin distributions \citep{hildebrand1983}. For face-on geometries, the IR luminosity first increases and then flattens out. However, for edge-on geometries, the IR luminosity increases and then decreases with increasing mass. This is also visible in the shape and flux densities of the higher-massed edge-on SEDs in Fig.\,\ref{largeseds}. The effect is that edge-on spirals and discs, that are optically thick, have fainter SEDs that also appear akin to cold BBs (see comparison with BBs in Sect.\,\ref{Sect:discussion}). This peculiar behaviour is caused by the fact that only the colder outer dust is observed in optically thick edge-on geometries. This complicates the $\chi_{\rm model}^2$-comparison with edge-on geometries. For these, the $C_M$ factors span an extended range even though the $\chi_{\rm model}^2$ values remain low for an extended mass range (as seen in Table\,\ref{largemassrelationtable}), making it difficult to constrain the $C_M$ factor.

\begin{figure}
    \centering
        \includegraphics[width=80mm]{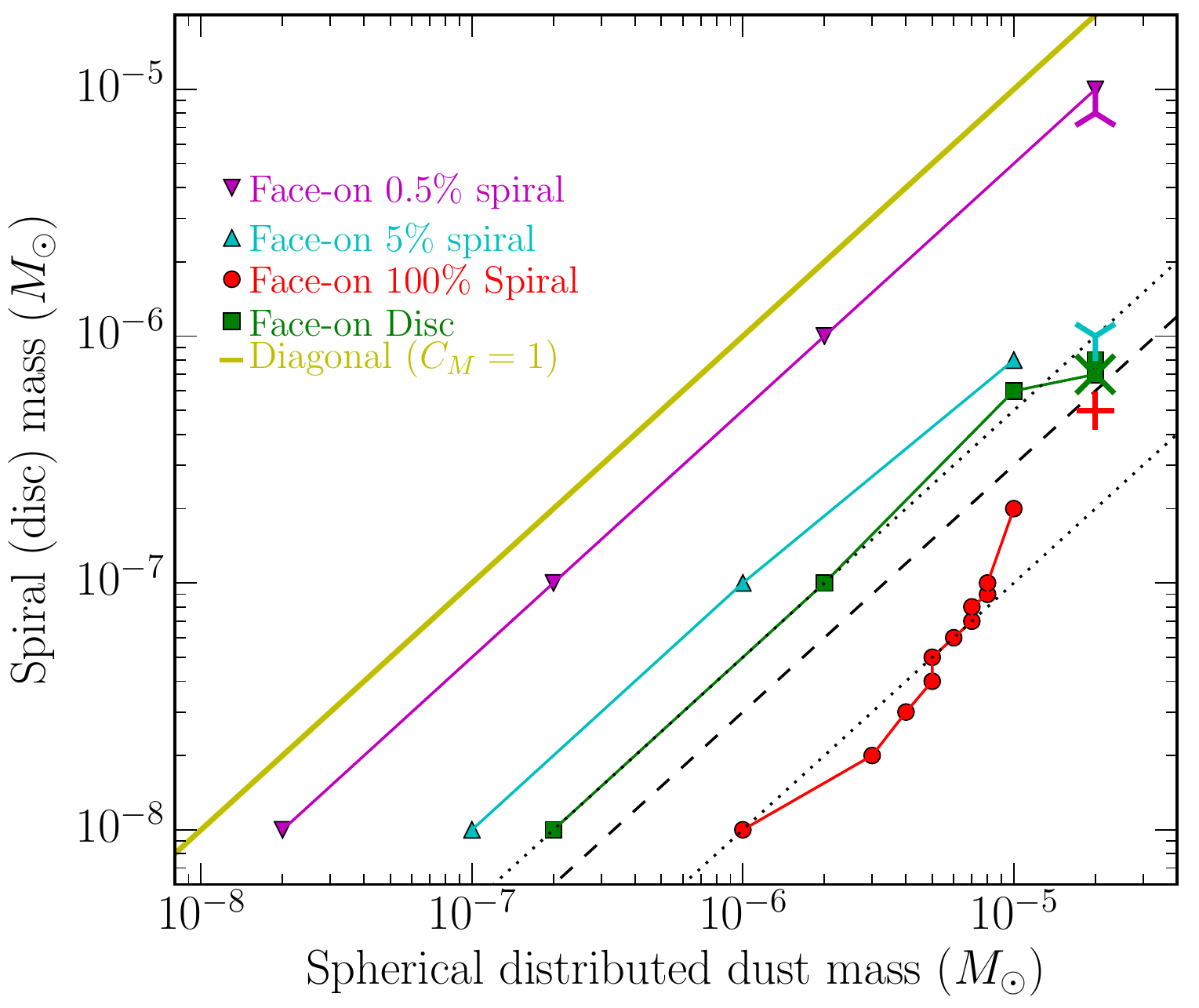}
    \caption{Masses of face-on distributions of spirals and discs on the y-axis, compared with corresponding spherical masses on the x-axis, from models Table\,\ref{largemassrelationtable}, i.e. SEDs of spirals and discs that are similar to sphere-SEDs according to the limit $\chi_{\rm model}^2~<~15$, and their corresponding masses. The black dashed line is a common mass correction for the non-spherical geometries, $C_M($combined$) \approx 0.03~\pm~0.02$, with the black dotted lines being the uncertainties. Mass corrections found with best fits to the \epaqr\ observed data (see Sect.~\ref{comparisonphotometry} for details) are included as a green cross (disc), red plus (100\%\ spiral), cyan \textit{Y} (5\%\ spiral), and magenta inverted \textit{Y} (0.5\%\ spiral). A diagonal yellow line is included to show where $C_M = 1$.}
        \label{largemassrelation}
\end{figure}

For the face-on geometries, the entire radial extent of the distributions are visible for observers, regardless of optical thickness, and thus the emission temperature changes much less when we compare different total dust masses. Therefore, we focus here on the $C_M$ factors of the face-on geometries and plot these in Fig.\,\ref{largemassrelation}. A caveat to keep in mind is the lack of silicate features for the included highest masses of the 100\%\ spirals, as shown in Figs.\,\ref{largeseds} and \ref{chisquare50}. This issue is connected to the aforementioned higher optical thickness for the spirals when compared to discs and spheres with the same dust mass.

In Fig.\,\ref{largemassrelation} we plot the face-on spiral and disc masses versus their corresponding sphere masses given in Table\,\ref{largemassrelationtable} with $\chi^2_{\rm model}~<~15$. For the face-on spirals without a spherical component we found an average $C_M($spiral$) = 0.011~\pm~0.003$, while for the discs we found an average $C_M($disc$) = 0.047~\pm~0.009$, where the errors are one standard deviation of the included corrections in Table\,\ref{largemassrelationtable}. We use these as the lower and higher limit when formulating the common correction, $C_M($combined$) \approx 0.03~\pm~0.02$, which is shown as black dashed and dotted lines in Fig.\,\ref{largemassrelation}. For the spiral-sphere combinations we found the corrections $C_M($5\%~spiral$) = 0.093~\pm~0.009$ and $C_M($0.5\%~spiral$) = 0.5~\pm~0.0$. As expected, decreasing the spiral components’ mass fraction increases the correction term and brings it closer to unity. The yellow line in Fig.\,\ref{largemassrelation} shows where $C_M = 1$, meaning the limit where the dust envelopes become purely spherical. These numbers show that by deducing a dust mass from an SED with an assumed spherical geometry, the real dust mass may be as low as 3\%\ of the deduced mass if the real geometry is a vertically flat spiral or disc.

We must stress the low number of data points here, and the large number of parameters that govern the distributions. For example, the mass corrections are only found within a small range of spherical-model dust masses, namely $1~\times~10^{-7}$ to $2~\times~10^{-5}$~M$_\odot$ for the 100\%\ spirals and discs, and we were not able to find SEDs similar enough to higher masses of spherical SEDs. All of these corrections are not generally applicable for a wider range of observed (dust sphere) masses.

\subsection{Impact of inclination}
\label{Sect:inclination}

We have so far compared geometries corresponding to face-on and edge-on inclination angles. For these cases we note that optically thick discs and spirals have significantly different SEDs when observed at the two different inclinations, mostly due to differences in optical thickness along the LOS. However, also the radiating surface is very different; the face-on surface is $\pi~\times~R_{\rm out}^2$, where $R_{\rm out}$ is the outer radius, while the edge-on surface is of the order of $2~R_{\rm out} \times h$, where $h$ is the distribution's vertical width. The emission flux density ratio between face-on and edge-on orientations is then proportional to the ratio of the surface areas, that is $\propto~R_{\rm out} / h$, which for a disc is as small as $\sim$0.003. From Fig.\ref{figspectralfeatures} we see that at 50~\um\ the flux density ratio is 0.01, which is consistent with this estimate (while at 10~\um\ we obtain a deep absorption).

Here we now compare the effects of different inclination angles. The spiral observed around \epaqr\ by \citet{homan2018} has an inclination between 4\adeg\ and 18\adeg . We used the most massive  ($10^{-4}$~M$_\odot$) spiral and disc models, with no spherical component, since these are the most optically thick and hence will have the most significant differences when comparing inclinations. We simulated SEDs at inclinations of 0\adeg , 25\adeg , 50\adeg , 60\adeg , 70\adeg , 80\adeg , and between 85\adeg\ and 90\adeg\ with increments of 1\adeg . Since we see significant impacts on the features at 10 and 18~\um , we compared the flux densities at 10 and 18~\um , and around the minima between these features, at 13~\um .

\begin{figure}
    \centering
        \includegraphics[width=80mm]{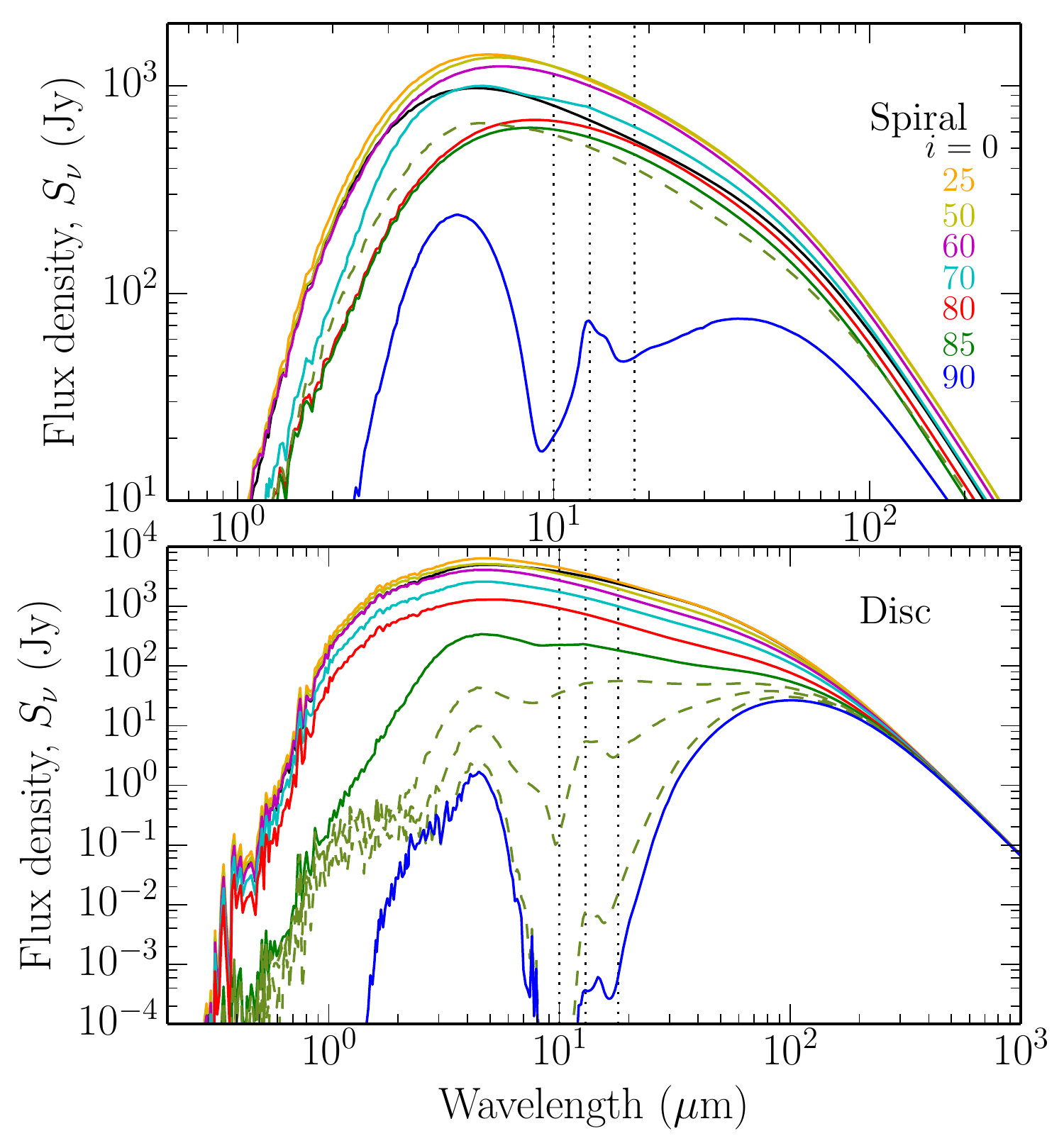}
    \caption{Simulated dust SEDs from spirals and discs without spherical components, and without including the stellar SED, at different inclinations. The green dashed lines are inclinations between 85\adeg\ and 90\adeg , where we show 89\adeg\ for the spiral and 87\adeg , 88\adeg , and 89\adeg\ for the disc.}
        \label{inclinationcompare}
\end{figure}

In Table\,\ref{inclinationcomparetable} we list the different flux ratios obtained from dust-only SEDs at the wavelengths 10, 13, and 18~\um , and in Fig.\,\ref{inclinationcompare} we plot the corresponding dust SEDs without the stellar SED. Here we can see that, with our choice of morphologies with limited scale heights, we do not obtain any significant impact for inclinations $i < 85$\adeg . For the dust disc, we find the most extreme cases of absorption features in the range 85\adeg -- 90\adeg , similar to what was found earlier. This connects to the choice of scale height and the lower optical thickness of the disc when compared to the spiral (see Table\,\ref{opticalthickness}), since a higher optical thickness extinguishes spectral lines. In any case, these results show that with our limited choice of spirals and disc geometries, they emit SEDs similar to those seen face-on for most inclinations.

\begin{table}
    \caption{List of flux ratios at wavelengths 10, 13, and 18~\um\ as a function of inclinations $i$.}
    \label{inclinationcomparetable}
    \begin{center}
    \begin{tabular}{rccc}
\hline\hline
\noalign{\smallskip}
    Spiral & $\frac{S(10\textrm{\um })}{S(13\textrm{\um })}$ & $\frac{S(18\textrm{\um })}{S(13\textrm{\um })}$ & $\frac{S(10\textrm{\um })}{S(18\textrm{\um })}$ \\
\noalign{\smallskip}
\hline
\noalign{\smallskip}
$i =$ 0\adeg & 1.19 & 0.80 & 1.49 \\
     25\adeg & 1.17 & 0.79 & 1.48 \\
     50\adeg & 1.15 & 0.79 & 1.45 \\
     60\adeg & 1.15 & 0.80 & 1.43 \\
     70\adeg & 1.10 & 0.80 & 1.36 \\
     80\adeg & 1.08 & 0.84 & 1.29 \\
     85\adeg & 1.09 & 0.83 & 1.32 \\
     86\adeg & 1.13 & 0.85 & 1.37 \\
     87\adeg & 1.15 & 0.87 & 1.41 \\
     88\adeg & 1.26 & 0.95 & 1.54 \\
     89\adeg & 1.29 & 0.92 & 1.54 \\
     90\adeg & 0.26 & 0.66 & 0.40 \\
\noalign{\smallskip}
\hline\hline
\noalign{\smallskip}
    Disc & $\frac{S(10\textrm{\um })}{S(13\textrm{\um })}$ & $\frac{S(18\textrm{\um })}{S(13\textrm{\um })}$ & $\frac{S(10\textrm{\um })}{S(18\textrm{\um })}$ \\
\noalign{\smallskip}
\hline
\noalign{\smallskip}
$i =$ 0\adeg & 1.21 & 0.75 & 1.62 \\
     25\adeg & 1.27 & 0.72 & 1.76 \\
     50\adeg & 1.27 & 0.70 & 1.81 \\
     60\adeg & 1.27 & 0.70 & 1.82 \\
     70\adeg & 1.26 & 0.71 & 1.77 \\
     80\adeg & 1.27 & 0.70 & 1.81 \\
     85\adeg & 0.98 & 0.79 & 1.24 \\
     86\adeg & 0.45 & 0.52 & 1.66 \\
     87\adeg & 0.15 & 0.24 & 1.94 \\
88\adeg &$5\times 10^{-4}$& 0.01            & 2.78 \\
89\adeg &$9\times 10^{-9}$&$7\times 10^{-5}$& 2.89 \\
90\adeg &$6\times 10^{-7}$& 1.70            &$3\times 10^{-7}$\\
\noalign{\smallskip}
\hline
    \end{tabular}
    \end{center}
\end{table}

\subsection{Comparison with observed photometry}
\label{comparisonphotometry}

To test the mass correction factor estimated in Sect.~\ref{Sect:masscorrection}, we also compare our models with the photometry of \epaqr\ presented in Table~\ref{allfluxes}, and with the  ISO-SWS and \textit{Herschel}-PACS spectra listed in Sect.~\ref{Sect:properties}. Our goal here is not to reproduce the observed dust emission to a high degree of accuracy, but to test the similarities and differences of these simulated SEDs to a realistic case. For studies that reproduce the dust emission, we refer to earlier publications, for example \citet{heras2005} who used one-dimensional radiative transfer for that purpose.

\begin{figure}
    \centering
        \includegraphics[width=80mm]{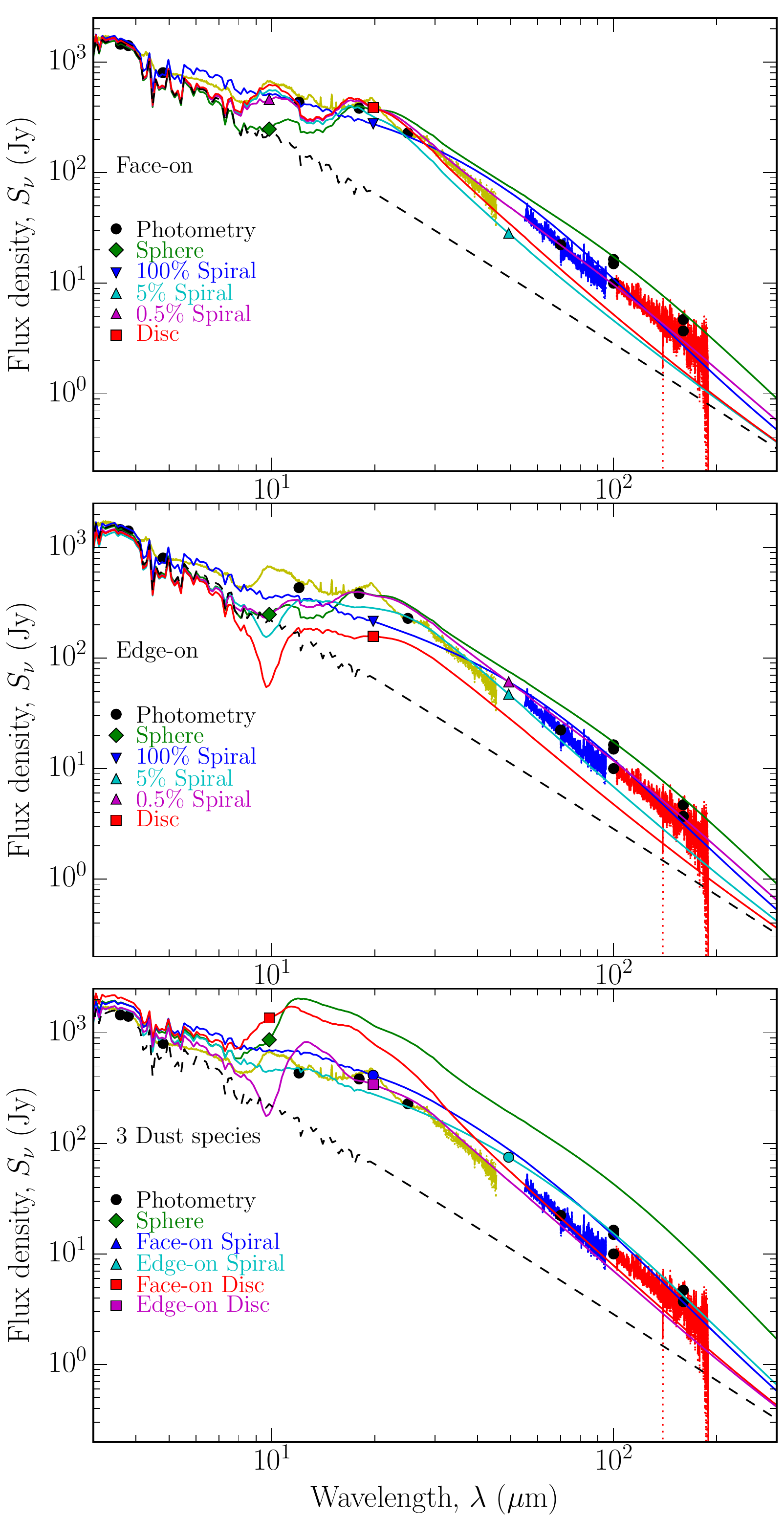}
    \caption{Comparison between best-fit models, photometry and IR spectra for \epaqr .  The \textit{top} and \textit{mid} panels use dust with 99\%\ Mg$_2$SiO$_4$ and 1\%\ Fe$_2$SiO$_4$. The \textit{bottom} panel compares the same dust masses and morphologies with these dust species: 90\%\ Mg$_2$SiO$_4$, 9\%\ Al$_2$O$_3$, and 1\%\ Fe$_2$SiO$_4$. The \textit{top} panel shows only SEDs from face-on geometry while the \textit{mid} panel shows only SEDs from an edge-on geometry. The different morphologies are indicated by different colours and symbols as annotated in the panels. The black dashed curve is the stellar photospheric SED model. The yellow spectrum is from ISO-SWS \citep{sloan2003}. The blue and red spectra are from \textit{Herschel}-PACS \citep{nicolaes2018}. The photometry (Table~\ref{allfluxes}) is shown as black dots (the error bars were smaller than the symbols and are not shown here).}
        \label{largedatacompare}
\end{figure}

We use the IR and FIR data, that is wavelengths longer than 2~\um\ where the relevant dust emission appear, and iterate through the different models to find the best-fitting one for each dust geometry with the help of Eq.~\ref{eqchidata}. The best fits are shown in Fig.~\ref{largedatacompare} and their corresponding dust masses and $\chi^2_{\rm red}$ values are listed in Table~\ref{largedatacomparetable}. The large  $\chi^2_{\rm red}$ values are due to the offset between the ISO-SWS spectrum and the \textit{Herschel}-PACS spectra.

\begin{table*}
    \caption{Best-fit dust models with two dust species (Mg$_2$SiO$_4$ and Fe$_2$SiO$_4$) and their $\chi^2_{\rm red}$ values for \epaqr . Smallest $\chi^2_{\rm red}$ for each data set are indicated with bold font.}
    \label{largedatacomparetable}
    \begin{center}
    \begin{tabular}{ccccccc}
\hline\hline
\noalign{\smallskip}
Morphology & Dust mass & Mass correction$^a$ & $\chi^2_{\rm red}$ & $\chi^2_{\rm red}$ & $\chi^2_{\rm red}$ & $\chi^2_{\rm red}$ \\
           & (M$_\odot$) & ($C_M$) & (photometry) & (ISO-SWS) & (PACS B) & (PACS R) \\
\noalign{\smallskip}
\hline
\noalign{\smallskip}
Sphere                   &$2\times 10^{-5}$& N/A  &    23.8 &     2581 &   1442   &   558   \\
Disc           (0\adeg ) &$7\times 10^{-7}$& 0.04 &    16.3 &{\bf 1049}&    552   &   231   \\
Disc          (90\adeg ) &$6\times 10^{-7}$& 0.03 &    22.3 &     6165 &    799   &   266   \\
5\%\ Spiral    (0\adeg ) &$1\times 10^{-6}$& 0.05 &    18.0 &     1187 &    872   &   275   \\
5\%\ Spiral   (90\adeg ) &$2\times 10^{-6}$& 0.01 &    12.2 &     3371 &    134   &   116   \\
0.5\%\ Spiral  (0\adeg ) &$8\times 10^{-6}$& 0.40 &     8.2 &     1250 &{\bf 28.9}&{\bf 8.5}\\
0.5\%\ Spiral (90\adeg ) &$1\times 10^{-5}$& 0.50 &     8.4 &     2539 &    168   &    38.7 \\
Spiral         (0\adeg ) &$5\times 10^{-7}$& 0.03 &     7.9 &     1141 &    344   &   13.5  \\
Spiral        (90\adeg ) &$6\times 10^{-7}$& 0.03 &{\bf 7.6}&     1838 &    335   &   30.7  \\
\noalign{\smallskip}
\hline
\noalign{\smallskip}
\multicolumn{3}{r}{\textit{Number of sampled points}} & 18 & 145 & 26 & 29 \\
    \end{tabular}
    \end{center}
    \begin{list}{}{}
    \item[Notes.]
The large $\chi^2$ values are due to (i) that our models are not tailored to fit the data (see text), and (ii) that there is an offset between ISO-SWS and \textit{Herschel}-PACS spectra.
$^a$ $C_M$ is the mass correction of Eq.\,\ref{eqmasscorr}.
    \end{list}
\end{table*}

In Fig.\,\ref{largedatacompare} we also include an additional panel with the same geometrical dust distributions, but with another dust composition: 90\%\ Mg$_2$SiO$_4$, 9\%\ Al$_2$O$_3$, and 1\%\ Fe$_2$SiO$_4$. We see here that the aluminium feature at $\sim 13$~\um\ dominates the spectrum. However, we see a similar trend as with the previous dust composition: Namely that the pure spiral, without a spherical component, has flatter and more featureless spectra than the other morphologies. The sphere and face-on disc spectra are most similar. When comparing the simulated spectra with the ISO-SWS spectrum, we can see that Al$_2$O$_3$ is not a significant ingredient of the dust of \epaqr .

When comparing the spectra in the two top panels in Fig.\,\ref{largedatacompare} (distributions of two-species), we find that the face-on disc and the sphere-spiral  distributions exhibit spectral features that correspond well to the ISO-SWS spectra, and the silicate features at 10 and 18~\um . However, the discs and 5\%\ spiral underestimate the \textit{Herschel}-PACS flux densities which indicate that they lack cold dust emission. For the edge-on discs and spirals it was not possible to find better fits since higher dust masses shift the SED-peak to longer wavelengths. In addition, models which are optically thick along the LOS do not fit the data either. However, this is consistent with \citet{heras2005} who found that the dust around \epaqr\ is optically thin. The best-fitting spherical model also exhibits the expected features at 10 and 18~\um ; however, it underestimates the flux density at 10~\um\ and marginally overestimates the \textit{Herschel}-PACS flux densities.

The two combined sphere-spiral geometries, when observed face-on, reproduce the features at 10 and 18~\um . The 0.5\%\ spiral also reproduces the FIR \textit{Herschel} spectrum whereas the 5\%\ spiral underestimates this spectrum. Not so surprising is the fact that the best-fit total dust masses of the sphere-spiral models are higher than for the 100\%\ spiral and disc, resulting in higher $C_M$ values than for those without a spherical component, as already seen in Fig.\,\ref{largemassrelation}. Additionally, the 100\%\ spiral model gives featureless SEDs at both inclination angles. This results in the face-on spiral fitting the photometry well but not the ISO-SWS spectrum.

To connect to the mass correction term $C_M$, we see in Table\,\ref{largedatacomparetable} that the 100\%\ face-on spiral and the disc result in $C_M = 0.03$ and 0.04, respectively, meaning that our spiral and disc masses, and the subsequent MLRs, are more than one order of magnitude smaller than that of the corresponding spherical distribution. These correspond well to what was found in Sect.~\ref{Sect:masscorrection}, that is $C_M($combined$) = $0.01 to 0.05. These were also plotted in Fig.\,\ref{largemassrelation} as cross and plus signs, respectively, for the spiral and disc. For the 5\%\ and 0.5\%\ sphere-spiral combinations, we find higher $C_M$ values than for the 100\%\ face-on spiral since these are similar to the spherical distribution, whereas the 0.5\%\ spiral has an order of magnitude larger $C_M$ value of 0.4. The sphere-spiral combined distributions are indicated in Fig.\,\ref{largemassrelation} as magenta and cyan {\it Y} symbols.

Our $C_M$ estimates from Sect.~\ref{Sect:masscorrection} are quite conservative. This is due to the combination of the large error in the mass correction, the many parameters (e.g. spiral and disc thickness, inter-arm distance, dust composition, stellar SED, sphere-spiral mass ratio), and the few data points we used to find $C_M$. However, we have shown that an observed SED may lead, when interpreted with a spherical model, to a total dust mass in a range that may be more than one order of magnitude larger than if the dust were distributed in a spiral or a disc. However, in the case of a realistic mass ratio between a dust spiral and the surrounding dust sphere (i.e. 5\%\ of the total dust mass residing in a spiral), the correction factor is $\sim 0.5$ instead. Depending on the ratio between the mass of the dust spiral and sphere, there should exist a range of correction factors between $\sim 0.01$ to 1, that is up to no difference at all.

\section{Conclusions} 

In this pilot study we used RADMC-3D to simulate SEDs for spherical, spiral, and disc-shaped dust distributions. The dust was distributed within the radial extensions of 5 to 5000~au around the O-rich AGB star \epaqr . We found significant differences when comparing SEDs from the more massive dust envelopes.

We found that the spirals and discs become partially optically thick at lower total dust masses than the equivalent sphere due to the higher local densities of the former. This is especially true when no spherical component was included. This causes differences in spectral features, mainly at the 10 and 18~\um\ silicate features. Optically thick discs and spirals, when seen face-on, have almost no strong spectral features and appear more akin to BBs. More massive edge-on discs exhibit an absorption at 10~\um\ which is overwhelmed by the stellar SED. The edge-on, non-spherical, high-mass geometries ($10^{-4}$ and $10^{-5}$~M$_\odot$) appear as BB at longer wavelengths. This is the opposite of how face-on distributions appear, since for edge-on distributions only the outer, colder dust is visible along the LOS, while with face-on and optically thick distributions, the warmer inner dust emission is still visible.

General conclusions from the spectral features are that i) if observations of an O-rich AGB star show significant warm dust emission, but no detectable silicate features, it would be prudent to consider non-spherical models; ii) if observations of an AGB star show colder CSE dust than expected, the cause could be an edge-on spiral or disc. However, the same effect may have other causes, for example larger dust grains.

We compared differences in dust masses of similar SEDs from spheres, face-on spirals and discs. We formulated a correction term which is the ratio between the mass of two different distributions with similar SEDs. This correction term  may be used to estimate the mass of a spiral or a disc even though a spherical model has been used to model the observed SED. In the extreme case of spirals and discs with no spherical component, we found that the correction term is between 0.01 to 0.05 times that of the mass of an assumed spherical distribution. This term is larger if a sphere is included around the spiral. For example, when the spiral consists of 5\%\ of the total dust mass, the mass correction is a factor of 0.09, and when the spiral is 0.5\% , the mass correction is 0.5.

These estimates were based on many assumptions (e.g. the disc and spiral morphologies), however, they are consistent with the masses of our dust geometries when they were compared with the observed far-IR photometric and spectroscopic data of \epaqr . The difference in mass of dust geometries that exhibit comparable SEDs may correspond to up to one to two orders of magnitude differences in estimated MLRs for an AGB star. However, additional comparisons of additional variations of dust morphology with additional known spiral-hosting AGB stars is required to constrain observable differences better. For example, it would be useful to study how a range of mass ratios between the spherical and spiral components of dust envelopes affects the correction term we formulated in Sect.\,\ref{Sect:masscorrection}.

\begin{acknowledgements}
This research has been funded by the Belgian Science Policy Office under contract BR/143/A2/STARLAB.
L. Decin acknowledges support from ERC consolidator grant 646758 AEROSOL.
We would like to thank the members of \textit{Team L.E.E.N.} at KU~Leuven for their interesting discussions, and general support for this project. We also want to thank Daniel Wiegert for his technical support with  significant computational time at his private cluster.
This work has made use of the SIMBAD database, operated at CDS, Strasbourg, France, the SVO Filter Profile Service supported from the Spanish MINECO through grant AYA2017-84089, and data from \textit{Herschel}-PACS. PACS has been developed by a consortium of institutes led by MPE (Germany) and including UVIE (Austria); KU~Leuven, CSL, IMEC (Belgium); CEA, LAM (France); MPIA (Germany); INAF-IFSI/OAA/OAP/OAT, LENS, SISSA (Italy); IAC (Spain). This development has been supported by the funding agencies BMVIT (Austria), ESA-PRODEX (Belgium), CEA/CNES (France), DLR (Germany), ASI/INAF (Italy), and CICYT/MCYT (Spain).
This work has also made use of data from the European Space Agency (ESA) mission {\it Gaia} (\url{https://www.cosmos.esa.int/gaia}), processed by the {\it Gaia} Data Processing and Analysis Consortium (DPAC, \url{https://www.cosmos.esa.int/web/gaia/dpac/consortium}). Funding for the DPAC has been provided by national institutions, in particular the institutions participating in the {\it Gaia} Multilateral Agreement.
\end{acknowledgements}


%
\bibliographystyle{aa}
\bibliography{referencesepaqr}
%

\end{document}